\documentclass[lettersize,journal]{IEEEtran}
\usepackage{amsmath,amsfonts}
\usepackage{algorithmic}
\usepackage{algorithm}
\usepackage{array}
\usepackage[caption=false,font=normalsize,labelfont=sf,textfont=sf]{subfig}
\usepackage{textcomp}
\usepackage{stfloats}
\usepackage{url}
\usepackage{verbatim}
\usepackage{graphicx}
\usepackage{color}
\usepackage{booktabs}
\usepackage{multirow}
\usepackage{algorithm}
\usepackage{algorithmic}
\usepackage{makecell}
\usepackage{amsmath}
\usepackage{ulem}
\UseRawInputEncoding
\begin{document}

\title{Quantum circuit for implementing AES S-box with low costs}

\author{
    \thanks{Manuscript created May, 2025. This research was funded by the National Natural Science Foundation of China (Grant Nos. 62171131, 62272056 and 62372048), Fujian Province Natural Science Foundation (Grant Nos. 2022J01186 and 2023J01533), Fujian Province Young and Middle-aged Teacher Education Research Project (Grant No. JAT231018)  and Open Foundation of State Key Laboratory of Networking and Switching Technology (Beijing University of Posts and Telecommunications) (SKLNST-2024-1-05).}
    Hui-Nan Chen\thanks{Hui-Nan Chen is with the College of Computer and Cyber Security, Fujian Normal University, Fuzhou 350117, China.},
    Bin-Bin Cai\thanks{Bin-Bin Cai is now with the College of Computer and Cyber Security, Fujian Normal University, Fuzhou 350117, China, the State Key Laboratory of Networking and Switching Technology, Beijing University of Posts and Telecommunications, Beijing 100876, China, and with the Digital Fujian Internet-of-Things Laboratory of Environmental Monitoring, Fujian Normal University, Fuzhou 350117, China (e-mail: cbb@fjnu.edu.cn).},
    Fei Gao\thanks{Fei Gao is now with the State Key Laboratory of Networking and Switching Technology, Beijing University of Posts and Telecommunications, Beijing 100876, China.},
    Song Lin\thanks{Song Lin is with the College of Computer and Cyber Security, Fujian Normal University, Fuzhou 350117, China (e-mail: lins95@fjnu.edu.cn).}
    \thanks{(Corresponding authors: Bin-Bin Cai, Song Lin.)}
}

\markboth{Journal of \LaTeX\ Class Files,~Vol.~14, No.~8, August~2021}%
{Shell \MakeLowercase{\textit{et al.}}: A Sample Article Using IEEEtran.cls for IEEE Journals}

\IEEEpubid{0000--0000/00\$00.00~\copyright~2021 IEEE}

\maketitle

\begin{abstract}
The Advanced Encryption Standard (AES) is widely used and well-studied for its efficiency and strong security. This paper presents quantum circuit designs for the AES S-box by introducing the composite field \( F((2^4)^2) \) to replace the traditional field \( F(2^8) \), enabling the inversion to be decomposed into operations over \( F(2^4) \). This work reduces the quantum resource overhead required for implementing the S-box by decreasing the number of $CNOT$ gates in the matrix multiplication, lowering the depth of $T$ gates in both the inversion circuit and the multiplication circuit. Besides, the widths for the S-box quantum circuits are also optimized during the inversion circuit and multiplication circuit. With a linear key schedule, the resulting AES-128 quantum circuit reduce the product of the circuit width and $T$ depth to 102800, which is the lowest known to date.

\end{abstract}


\section{Introduction}

\IEEEPARstart{I}{n} recent years, the deep integration of quantum mechanics and information science has driven the rapid advancement of quantum information \cite{r1,r2}, positioning it as a frontier field in scientific research and technological innovation. As a vital branch of this field, quantum computing has attracted widespread attention due to its unique computational model and powerful parallel processing capabilities. Unlike traditional computers that rely on binary bit states, quantum computers leverage the principles of superposition and entanglement, allowing qubits to process multiple states simultaneously and achieve unprecedented parallel computing capabilities \cite{r3,r4,r5,r6}. This grants quantum computing the potential to surpass classical computation, especially by offering significant speedup in solving certain complex problems \cite{r7,r8,r9}. In recent years, ongoing progress in areas, such as quantum architecture search \cite{r10}, variational quantum algorithm \cite{r11}, and quantum neural network \cite{r12}, has significantly broadened the scope of quantum computing in learning systems and secure information processing. This paradigm shift has profoundly impacted various disciplines, among which cryptography is one of the most significantly affected.

 In the evaluation of cryptographic system security, quantum algorithms offer unprecedented acceleration in the attack process. Representative examples include Shor＊s algorithm \cite{r13,r14}, Grover＊s algorithm \cite{r15,r16}, and Simon＊s algorithm \cite{r17,r18}. Shor's algorithm leverages quantum computation to \IEEEpubidadjcol determine the periodicity of a function and employs the quantum Fourier transform \cite{r19,r20,r21} to efficiently solve the factoring problem, significantly accelerating the decomposition of large integers. Compared with the exponential complexity of classical algorithms, Shor's algorithm completes this task in polynomial time. As a result, it poses a direct threat to classical encryption systems based on the hardness of integer factorization, particularly the widely used RSA \cite{r22}. Grover＊s algorithm does not pose a direct threat to public-key cryptosystems as Shor＊s algorithm, while it has significant implications for the security of symmetric encryption algorithms, such as AES \cite{r23}. In the classical computing environment, performing an exhaustive key search on a symmetric cipher with an \( n \)-bit key typically requires \( O(2^n) \) operations. However, when Grover＊s algorithm is applied in a quantum computing context, the complexity is reduced to \( O(2^{n/2}) \), achieving a quadratic speedup. Simon＊s algorithm is one of the earliest quantum algorithms to demonstrate exponential speedup over classical counterparts. In recent years, it has shown strong capability in identifying hidden periodic structures within symmetric cryptographic constructions, such as the Even每Mansour cipher \cite{r24} and Feistel networks \cite{r25}. Although these theoretical advancements demonstrate the immense potential of quantum computing in cryptanalysis, a critical challenge remains in practical attacks, that is, how to design and implement the quantum $Oracle$ which is tailored to specific encryption algorithms. Therefore, constructing an efficient and feasible quantum $Oracle$ has become a key challenge in advancing quantum cryptanalysis from theory to practice.

 As a core component of quantum algorithms, the $Oracle$ serves to encapsulate the target function into a quantum subcircuit that can be queried by the algorithm, enabling effective access to and manipulation of the internal structure of cryptographic schemes. However, designing an $Oracle$ is often highly complex, as it requires faithfully replicating the classical function＊s logic while also carefully managing key quantum resources, such as the depth of the quantum circuit and the number of qubits \cite{r42,r33}. Therefore, minimizing the resource overhead of quantum circuit implementations while maintaining functional correctness has become one of the central challenges in quantum cryptanalysis \cite{r43}. Resource estimation and optimization around $Oracle$ construction have emerged as critical steps toward the practical realization of quantum attack strategies.

 As one of the most widely used symmetric encryption standards, AES has naturally become a core target in quantum cryptanalysis. Quantum resources required by Grover＊s algorithm to attack AES have been used as a reference benchmark for evaluating the security strength of post-quantum cryptographic algorithms in 2016 post-quantum standardization call of the National Institute of Standards and Technology (NIST) \cite{r29,r30}. Therefore, constructing optimized quantum circuits for AES is of critical importance for accurately assessing the resource requirements of Grover search attack. To achieve such optimization, it is necessary to implement a quantum $Oracle$ for AES and perform quantum resource estimation and refinement. Previous works on AES quantum circuit design have primarily focused on three directions: reducing the width (i.e., the number of qubits), the depth, or the product of width and depth.

 To reduce the quantum resource requirements of Grover search attack, researchers have focused primarily on reducing the width of the quantum circuit of AES. In 2016, Grassl et al. \cite{r31} designed a quantum circuit for the S-box with a width of 40, using a zig-zag approach. They constructed a quantum circuit for AES-128 with a final circuit width of 984. In 2018, Almazrooie et al. \cite{r32} optimized the key expansion strategy, further reducing the width of AES-128 to 976. In 2020, Langenberg et al. \cite{r33} proposed a quantum circuit for the S-box with a width of 32, incorporating a zig-zag structure in the key expansion, which led to a reduction in the width of AES-128 to 864. In the same year, Zou et al. \cite{r34} designed a quantum circuit for the S-box with a width of 22 and combined it with an improved zig-zag method with a width of 23, leading to the implementation of quantum circuit for AES-128 with a width of 512. In 2022, Wang et al. \cite{r35} presented a straight-line iterative method for key expansion, reducing the width of AES-128 to 400. Later, Li et al. \cite{r36} further minimized the width of AES-128, reducing it to 270 by constructing the quantum circuit for the S-box with a width of 22 and applying a streamlining optimization strategy. In 2023, Li et al. \cite{rr37} further improved the previous quantum circuits, reducing the width of AES-128 to 264. In 2025, Huang et al. \cite{r37} proposed a quantum circuit for AES-128 that requires only 256 qubits and does not rely on any additional ancillary qubits.

 In addition to width, depth of the quantum circuit for AES also plays a crucial role in determining the feasibility of quantum attacks. Therefore, considerable effort has been devoted to optimizing the sequence of gate operations to a lower depth. Jang et al. \cite{rr38} proposed the first S-box quantum circuit designed to reduce the depth. In their implementation, the AES-128 quantum circuit achieved a $T$-depth of 120. Huang and Sun \cite{rr39}, based on the classical circuit for the S-box proposed by Boyar and Peralta \cite{rr40}, constructed a quantum circuit of the S-box focused on optimizing the depth of $T$ gates. Their design reduced the $T$-depth of the AES-128 quantum circuit to 60.

 Furthermore, the cost of quantum attacks is also evaluated by the product of width and depth \cite{r42}. The metric was first proposed by Jaques et al. \cite{r42} and refined by Jang et al. \cite{rr38}. In 2023, Liu et al. \cite{rr41} proposed a technique called m-XOR to identify reusable qubits. They also designed a compact quantum circuit for S-box and ultimately constructed an AES-128 quantum circuit with the DW($T$) value of 147560. In 2025, Jiang et al. \cite{rr42} applied the controlled control qubit cascade (CCQC) technique to quantum circuit construction and built a quantum circuit for AES-128 with the DW($T$) value of 128640.

 \noindent\textbf{Contribution} In this paper, we refine the method proposed in Ref.\cite{r36} and develop a quantum implementation of AES with lower resource overhead, reducing the the product of the quantum circuit width and T depth (i.e., DW($T$)). The approach analyzes the algebraic structure of the S-box and introduces the composite field \( F((2^4)^2) \), which is isomorphic to the finite field \( F(2^8) \). This transformation allows the inversion of elements in \( F(2^8) \) to be performed within \( F((2^4)^2) \). Furthermore, the inversion in \( F((2^4)^2) \) is broken down into a sequence of operations over the smaller finite field \( F(2^4) \), which facilitates a quantum circuit design for the S-box with the reduced DW($T$) value. In the process mentioned above, we optimize the quantum implementation of matrix multiplication in \( F(2^4) \) by reducing the number of $CNOT$ gates, lowering the number of $T$ gates and $T$ depth in the inversion circuit in \( F(2^4) \), and achieving the quantum multiplication circuit in \( F(2^4) \) with the reduced width and depth. Therefore, we obtain the currently lowest known DW($T$) value for AES-128, which is 102800. Our design shows significant advantages under the MAXDEPTH parameter, proposed by NIST in 2016 \cite{r29,r30} to evaluate quantum circuit depth and attack cost. Compared with the work in Ref.\cite{rr42}, the DW($T$) value in our quantum circuit of AES-128 is reduced by 20.09\%.

 \section{Preliminaries}
 In this section, we provide an overview of the AES block cipher, composite field arithmetic, and some basic quantum gates.

 \subsection{The AES block cipher}
 AES uses a 128-bit block size and supports key lengths of 128, 192, and 256 bits, known as AES-128, AES-192, and AES-256, respectively. The number of encryption rounds $N_R$ for AES is 10, 12, and 14 for each key length. The plaintext $M$ undergoes one AddRoundKey operation, followed by $N_R$ rounds of the round function, ultimately producing the ciphertext $C$. Each round function consists of SubBytes, ShiftRows, MixColumns and AddRoundKey, except for the final round function, which contains only SubBytes, ShiftRows, and AddRoundKey.

 \subsubsection{The S-box} \label{Sect2.1.1}
 SubBytes, which is denoted as $S(a)$, represents the only nonlinear transformation within AES. It is performed by applying the same S-box transformation to each byte individually. The S-box is an $8\times8$ lookup table that maps an input byte to an output byte. $S(a)$ takes a byte $a\in F(2^8)$ as input, where $a=(a_7\ a_6\ a_5\ a_4\ a_3\ a_2\ a_1\ a_0)$ can be represented by the polynomial $(a_7x^7+a_6x^6+a_5x^5+a_4x^4+a_3x^3+a_2x^2+a_1x+a_0)$ with coefficients $\{0,1\}$.

 To implement the quantum circuit implementation of the S-box, we primarily focus on its algebraic representation. The process begins by computing the multiplicative inverse of $a$, denoted as $a^{-1}$, followed by an affine transformation that involves matrix multiplication and modular addition with a constant vector. Consequently, the algebraic form of $S(a)$ can be expressed as
\begin{equation}
 \label{ali1}
 S(a)=Aa^{-1}\oplus c,
\end{equation}
 where the matric $A$ and vector $c$ are defined as
 \begin{equation*}
A=
\begin{pmatrix}
  1 & 0 & 0 & 0 & 1 & 1 & 1 & 1\\
  1 & 1 & 0 & 0 & 0 & 1 & 1 & 1\\
  1 & 1 & 1 & 0 & 0 & 0 & 1 & 1\\
  1 & 1 & 1 & 1 & 0 & 0 & 0 & 1\\
  1 & 1 & 1 & 1 & 1 & 0 & 0 & 0\\
  0 & 1 & 1 & 1 & 1 & 1 & 0 & 0\\
  0 & 0 & 1 & 1 & 1 & 1 & 1 & 0\\
  0 & 0 & 0 & 1 & 1 & 1 & 1 & 1
\end{pmatrix}
,\ c=
\begin{pmatrix}
  1\\
  1\\
  0\\
  0\\
  0\\
  1\\
  1\\
  0
\end{pmatrix}
.
\end{equation*}

\subsubsection{The key expansion process} \label{Sect2.1.2}
AES-128, AES-192, and AES-256 have different key expansion processes due to their varying initial key lengths. Considering 32 bits as a word, the initial key $K_0$ can be divided into $N_K$ words for AES-128, AES-192 and AES-256, where $N_K$ equals 4, 6 and 8. For AES-128, the initial key can be represented as $K_0=W_3W_2W_1W_0$. After $N_R$ round key expansion process, $(N_R+1)$ subkeys are generated. The subkey for the $i$-th round of AES-128, AES-192, and AES-256 are all denoted as $K_i=W_{4i+3}W_{4i+2}W_{4i+1}W_{4i}$, where $0 \leq i \leq N_R$. Each subkey is 128 bits in length. The key expansion process for AES-128 and AES-192 is shown in Algorithm \ref{Alg1}, while the key expansion process for AES-256 is shown in Algorithm \ref{Alg2}. In these algorithms, $RotWord$ represents a circular left shift of a word by 8 bits, and $Rcon[j]$ denotes that $XOR$ the current state with a round constant, where $j \geq 0$. Each round constant $Rcon[j]$ is defined as a four-byte word of the form $(RC[j], 0x00, 0x00, 0x00)$, where only the first byte $RC[j]$ is nonzero and the remaining three bytes are zero. The value of $RC[j]$ corresponds to the element $x^{j-1}$ in the finite field $F(2^8)$, reduced modulo the irreducible polynomial $m(x) = x^8 + x^4 + x^3 + x + 1$, which serves as the generator polynomial in AES.

\begin{algorithm}
    \caption{The key expansion process for AES-128 and AES-192}
    \label{Alg1}
    \begin{algorithmic}
        \FOR{$t$ from $N_K$ to $4 \times (N_R+1)$}
            \IF{$t \bmod 4 == 0$}
                \STATE $W_t = W_{t-4} \oplus SubBytes(RotWord(W_{t-1})) \oplus Rcon[t/4]$
            \ELSE
                \STATE $W_t = W_{t-4} \oplus W_{t-1}$
            \ENDIF
        \ENDFOR
    \end{algorithmic}
\end{algorithm}

\begin{algorithm}
    \caption{The key expansion process for AES-256}
    \label{Alg2}
    \begin{algorithmic}
        \FOR{$t$ from $N_K$ to $4 \times (N_R+1)$}
            \IF{$t \bmod 8 == 0$}
                \STATE $W_t = W_{t-8} \oplus SubBytes(RotWord(W_{t-1})) \oplus Rcon[t/8]$
            \ELSIF{$t \bmod 8 == 4$}
                \STATE $W_t = W_{t-8} \oplus SubBytes(RotWord(W_{t-1}))$
            \ELSE
                \STATE $W_t = W_{t-8} \oplus W_{t-1}$
            \ENDIF
        \ENDFOR
    \end{algorithmic}
\end{algorithm}

\subsection{Composite field arithmetic} \label{Sect2.2}
In combinatorial domain operations, elements in higher-order domains can be expressed as linear combinations of elements in lower-order domains, where operations in lower-order domains are simpler and less costly. Therefore, solving the multiplicative inverse of any element $a$ in a finite field can be transformed into solving the multiplicative inverse of the corresponding element in a combinatorial domain using a mapping matrix $M$. After obtaining the multiplicative inverse of the corresponding element in the combinatorial domain, the multiplicative inverse of $a$ can be derived using the inverse matrix $M^{-1}$, resulting in $a^{-1}$.

Wolkerstorfer et al. \cite{r38} proposed an implementation of the S-box using the composite fields $F((2^4)^2)$ and $F(2^4)$, i.e.,\begin{equation*}
\begin{cases}
F((2^4)^2):  x^2 + x + \lambda, \\
F(2^4):  x^4 + x + 1,
\end{cases}
\end{equation*}
where
\begin{equation}
 \label{ali2}
 \lambda = x^3+x^2+x \in F(2^4).
\end{equation}
The mapping matrix $M:\ F(2^8) \to F((2^4)^2)$ and the inverse matrix $M^{-1}:\ F((2^4)^2) \to F(2^8)$ are defined as
\begin{equation*}
M=
\begin{pmatrix}
  1 & 0 & 0 & 0 & 1 & 1 & 1 & 0\\
  0 & 1 & 1 & 0 & 0 & 0 & 0 & 0\\
  0 & 1 & 0 & 0 & 0 & 0 & 0 & 1\\
  0 & 0 & 1 & 0 & 1 & 0 & 0 & 0\\
  0 & 0 & 0 & 0 & 1 & 1 & 1 & 0\\
  0 & 1 & 0 & 0 & 1 & 0 & 1 & 1\\
  0 & 0 & 1 & 1 & 0 & 1 & 0 & 1\\
  0 & 0 & 0 & 0 & 0 & 1 & 0 & 1
\end{pmatrix},
\end{equation*}
\begin{equation*}
M^{-1}=
\begin{pmatrix}
  1 & 0 & 0 & 0 & 1 & 0 & 0 & 0\\
  0 & 0 & 0 & 0 & 1 & 1 & 0 & 1\\
  0 & 1 & 0 & 0 & 1 & 1 & 0 & 1\\
  0 & 1 & 0 & 0 & 1 & 1 & 1 & 0\\
  0 & 1 & 0 & 1 & 1 & 1 & 0 & 1\\
  0 & 0 & 1 & 0 & 1 & 1 & 0 & 0\\
  0 & 1 & 1 & 1 & 1 & 0 & 0 & 1\\
  0 & 0 & 1 & 0 & 1 & 1 & 0 & 1
\end{pmatrix}.
\end{equation*}

It follows that the algebraic expression of Eq.(\ref{ali1}) can be rewritten as
 \begin{equation}
 \label{ali3}
 S(x)=AM^{-1}(Ma)^{-1} \oplus c,
\end{equation}
where
\begin{equation*}
    AM^{-1}=
\begin{pmatrix}
  1 & 0 & 1 & 0 & 1 & 1 & 0 & 1\\
  1 & 1 & 1 & 1 & 1 & 1 & 0 & 1\\
  1 & 0 & 0 & 1 & 1 & 1 & 0 & 0\\
  1 & 0 & 1 & 0 & 1 & 0 & 1 & 1\\
  1 & 1 & 0 & 1 & 1 & 0 & 1 & 1\\
  0 & 1 & 1 & 1 & 1 & 1 & 1 & 1\\
  0 & 0 & 0 & 0 & 1 & 0 & 1 & 1\\
  0 & 1 & 1 & 0 & 1 & 0 & 1 & 1
\end{pmatrix}
.
\end{equation*}

From Eq.(\ref{ali3}), it can be seen that $S(a)$ still requires the affine transformation, but now the multiplicative inverse is computed for the element in $F((2^4)^2)$, rather than in $F(2^8)$.

Any element in $F((2^4)^2)$ can be expressed as $p=p_0+p_1x$, where $p_0,\ p_1\in F(2^4)$. Ref.\cite{r38} shows that $p^{-1}$ can be expressed as
 \begin{equation}
 \label{ali4}
 p^{-1}=(p^{17})^{-1}(p_0+p_1)+(p^{17})^{-1}p_1x,
\end{equation}
 where
 \begin{equation}
\label{ali5}
 p^{17}={p_1}^2 \times \lambda +(p_0+p_1)p_0.
\end{equation}

From Eq.(\ref{ali4}) and Eq.(\ref{ali5}), $p^{-1}$ can be computed by $(p^{17})^{-1}(p_0+p_1)$, $(p^{17})^{-1}p_1x$, ${p_1}^2 \times \lambda$, and $(p_0+p_1)p_0$.

\subsection{Basic quantum gates}

\begin{figure}[h!]
    \centering
    \includegraphics[width=0.5\textwidth]{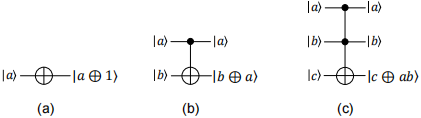}
    \caption{Some basic quantum gates used in this paper. (a) The $NOT$ gate; (b) The $CNOT$ gate; (c) The $Toffoli$ gate.}
    \label{fig1}
\end{figure}

Figure \ref{fig1} illustrates some basic quantum gates used in this paper. In particular, for the $CNOT$ gate (see Figure \ref{fig1}(b)) and the $Toffoli$ gate (see Figure \ref{fig1}(c)), the last qubit is the target qubits, while the remaining qubits are the control qubits.

\begin{figure}[h!]
    \centering
    \includegraphics[width=0.45\textwidth]{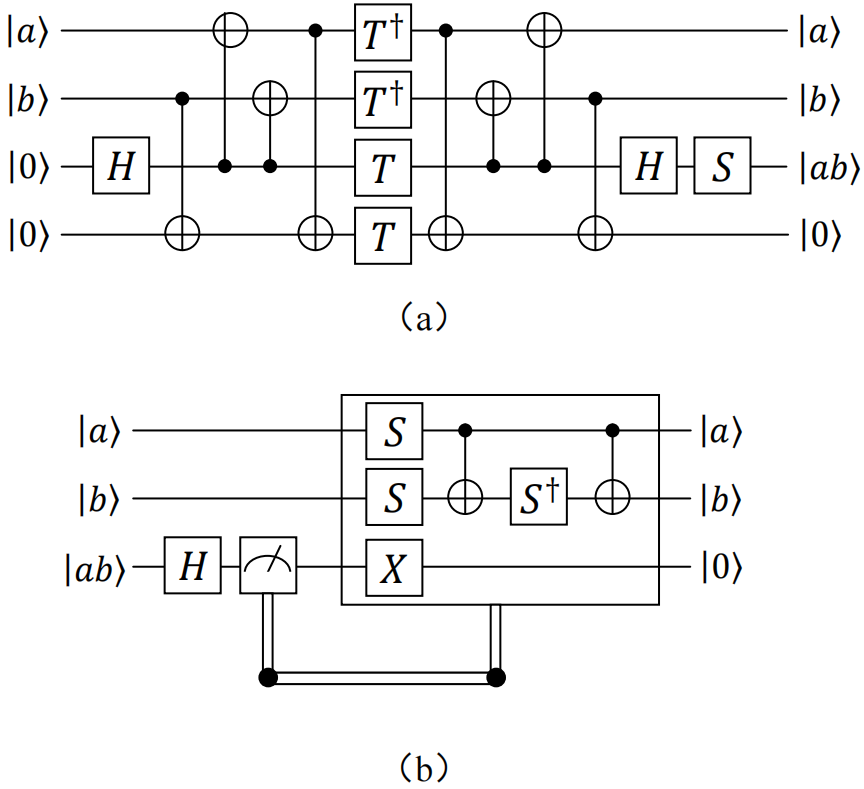}
    \caption{The $QAND$ gate and ${QAND}^\dagger$ gate \cite{r42}. (a) The $QAND$ gate. (b) The ${QAND}^\dagger$ gate.}
    \label{fig18}
\end{figure}

To compare with other literatures, the $QAND$ and $QAND^\dagger$ gates \cite{r42} shown in Figure \ref{fig18} are used in this paper to replace $Toffoli$ gates, thereby reducing the number of $T$ gates and the $T$ depth.

\section{The quantum circuits of the S-box}
The SubBytes operation is the sole nonlinear component of AES. Optimizing its quantum resource requirements is essential to achieve a cost-effective implementation of the quantum circuit for AES. This section leverages the algebraic structure of the S-box (Section \ref{Sect2.1.1}) and composite field arithmetic (Section \ref{Sect2.2}) to design efficient quantum circuits and compares the results with existing literatures.

\subsection{Quantum circuit implementations of $M$ and $AM^{-1}$} \label{Sect3.1}
A commonly method for implementing quantum circuits for matrix multiplication is the Permutation-Lower-Upper (PLU) decomposition. However, this approach often leads to a significant increase in the number of $CNOT$ gates. Xiang et al. \cite{r40} propose a heuristic algorithm for optimizing matrix implementation based on matrix decomposition theory, which offers advantages in reducing the number of $CNOT$ gates required.

\begin{figure}[h!]
    \centering
    \includegraphics[width=0.45\textwidth]{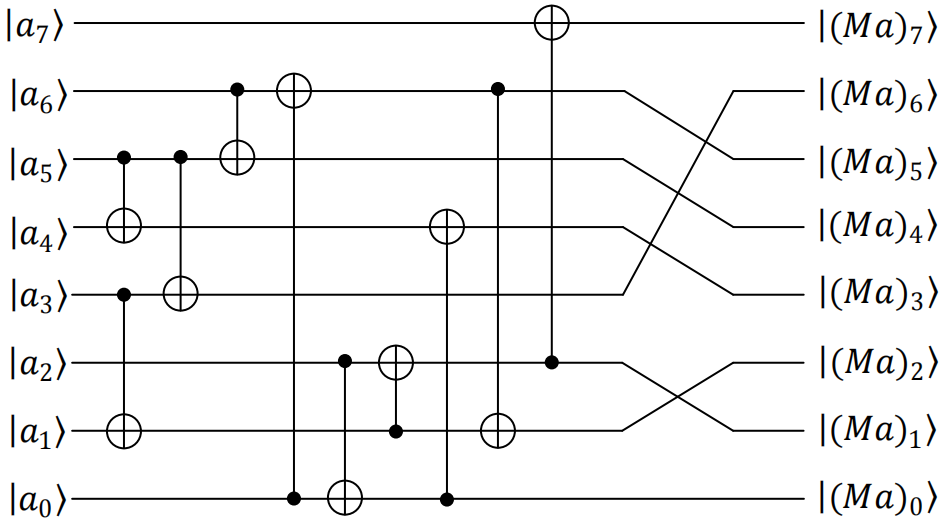}
    \caption{The quantum circuit for $U_M: |a\rangle \to |Ma\rangle$.}
    \label{fig2}
\end{figure}

\begin{figure}[h!]
    \centering
    \includegraphics[width=0.45\textwidth]{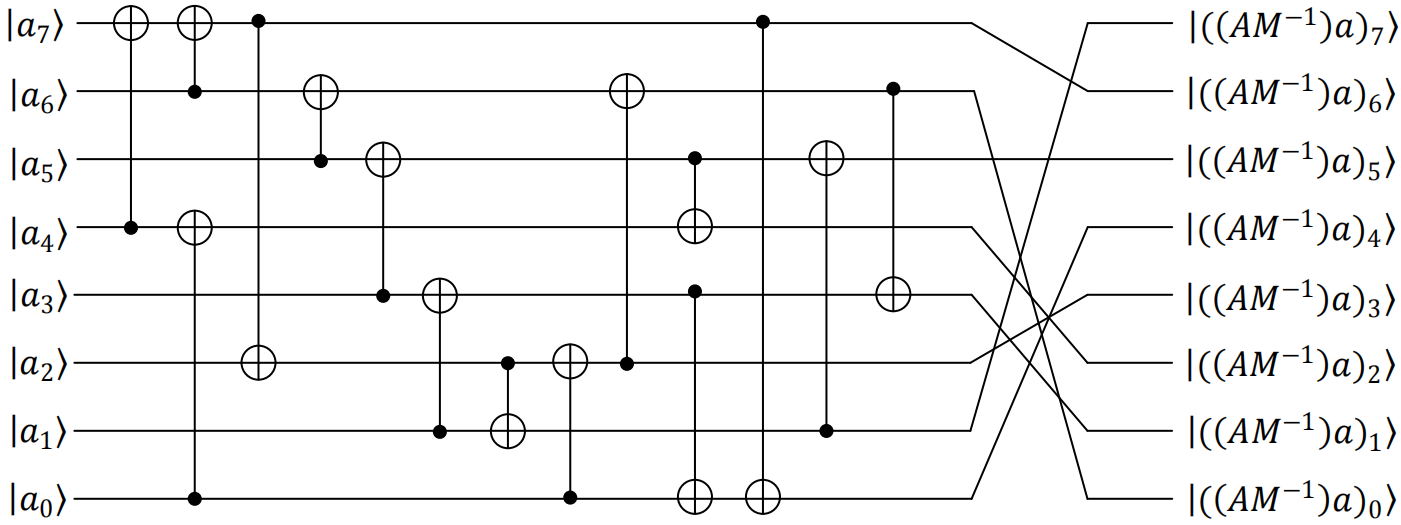}
    \caption{The quantum circuit for $U_{AM^{-1}}: |a\rangle \to |(AM^{-1})a\rangle$.}
    \label{fig3}
\end{figure}

In this paper, we apply the algorithm from Ref.\cite{r40} to implement the quantum circuits for matrix multiplication of $U_M: |a\rangle \to |Ma\rangle$ and $U_{AM^{-1}}: |a\rangle \to |(AM^{-1})a\rangle$, as shown in Figure \ref{fig2} and Figure \ref{fig3}, where $a=(a_7\ a_6\ a_5\ a_4\ a_3\ a_2\ a_1\ a_0)$. $U_M$ can be implemented with 8 qubits and 10 $CNOT$ gates, while $U_{AM^{-1}}$ can be implemented with 8 qubits and 15 $CNOT$ gates.

\subsection{Quantum circuit implementations of the multiplicative inversion in $F(2^4)$} \label{sect3.2}

Ref.\cite{r36} considers solving the multiplicative inverse of an element in $F(2^4)$ as a $4 \times 4$ S-box. Let $\boldsymbol{b}$ be any element in $F(2^4)$, with its corresponding multiplicative inverse $\boldsymbol{b^{-1}}$ shown in Table \ref{tab1}.
\begin{table}[!t]
\caption{The input elements and corresponding inverse output elements in $F(2^4)$\label{tab1}}
\centering
\scriptsize
\renewcommand{\arraystretch}{1.3}
\setlength{\tabcolsep}{4pt}
\begin{tabular}{|c|*{16}{c|}}
\hline
$\boldsymbol{b}$ & 0 & 1 & 2 & 3 & 4 & 5 & 6 & 7 & 8 & 9 & A & B & C & D & E & F \\
\hline
$\boldsymbol{b^{-1}}$ & 0 & 1 & 9 & E & D & B & 7 & 6 & F & 2 & C & 5 & A & 4 & 3 & 8 \\
\hline
\end{tabular}
\end{table}

\begin{figure}[h!]
    \centering
    \includegraphics[width=0.45\textwidth]{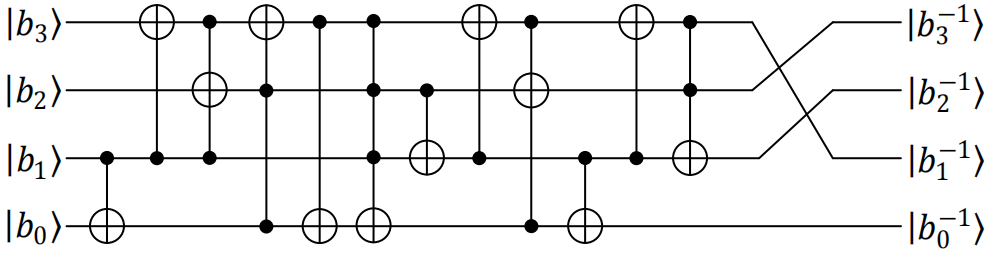}
    \caption{The quantum circuit for ${F(2^4)}_{inv}:\ |\boldsymbol{b}\rangle \to |\boldsymbol{b^{-1}}\rangle$.}
    \label{fig4}
\end{figure}

In this paper, we use the automated tool DORCIS \cite{r41} to realize the quantum circuit of the $4 \times 4$ lookup table, which solves the multiplicative inverse of elements in $F(2^4)$. The resulting quantum circuit is denoted as ${F(2^4)}_{inv}:\ |\boldsymbol{b}\rangle \to |\boldsymbol{b^{-1}}\rangle$, where $\boldsymbol{b}=(b_3\ b_2\ b_1\ b_0)$ is the input element and $\boldsymbol{b^{-1}}=(b^{-1}_3\ b^{-1}_2\ b^{-1}_1\ b^{-1}_0)$ is the output element, as shown in Figure \ref{fig4}.

\begin{figure}[h!]
    \centering
    \includegraphics[width=0.35\textwidth]{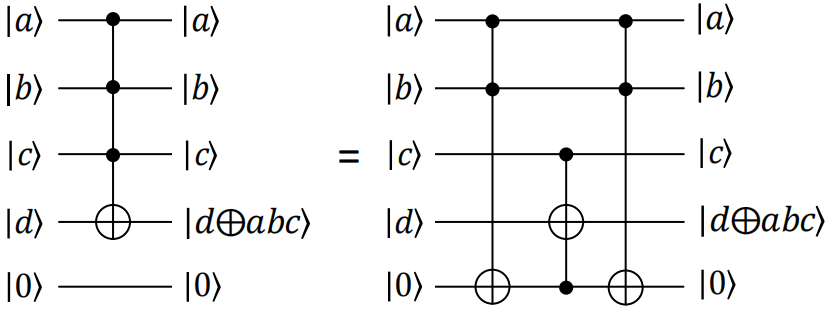}
    \caption{The decomposition of $3-controlled$ $NOT$ gate with a clean auxiliary qubit.}
    \label{fig5}
\end{figure}

The $3-controlled$ $NOT$ gate appearing in ${F(2^4)}_{inv}$ (shown in Figure \ref{fig4}) can be implemented using the decomposition method $|a\rangle|b\rangle|c\rangle|d\rangle|0\rangle \to |a\rangle|b\rangle|c\rangle|d \oplus abc\rangle|0\rangle$ (shown in Figure \ref{fig5}), which requires a clean auxiliary qubit, resulting in the decomposed circuit ${F(2^4)}_{inv0}:\ |\boldsymbol{b}\rangle|0\rangle \to |\boldsymbol{b^{-1}}\rangle|0\rangle$.

Furthermore, to further reduce the depth of $Toffoli$ gates, we derive the algebraic expression of ${F(2^4)}_{inv}$ (shown in Figure \ref{fig4}), i.e., \\
\begin{equation}
\label{ali6-1}
\begin{split}
{b_3}^{-1} &= b_3 + b_2 + b_1 + b_3 b_2 + b_3 b_1 + b_3 b_0 + b_3 b_2 b_1, \\
{b_2}^{-1} &= b_3 + b_2 + b_3 b_0 + b_2 b_0 + b_1 b_0 + b_3 b_2 b_0, \\
{b_1}^{-1} &= b_3 + b_3 b_1 + b_2 b_1 + b_2 b_0 + b_1 b_0 + b_3 b_1 b_0, \\
{b_0}^{-1} &= b_3 + b_2 + b_1 + b_0 + b_2 b_1 + b_2 b_0 + b_3 b_2 b_1 + b_2 b_1 b_0.
\end{split}
\end{equation}

\begin{figure*}[!t]
    \centering
    \includegraphics[width=1\textwidth]{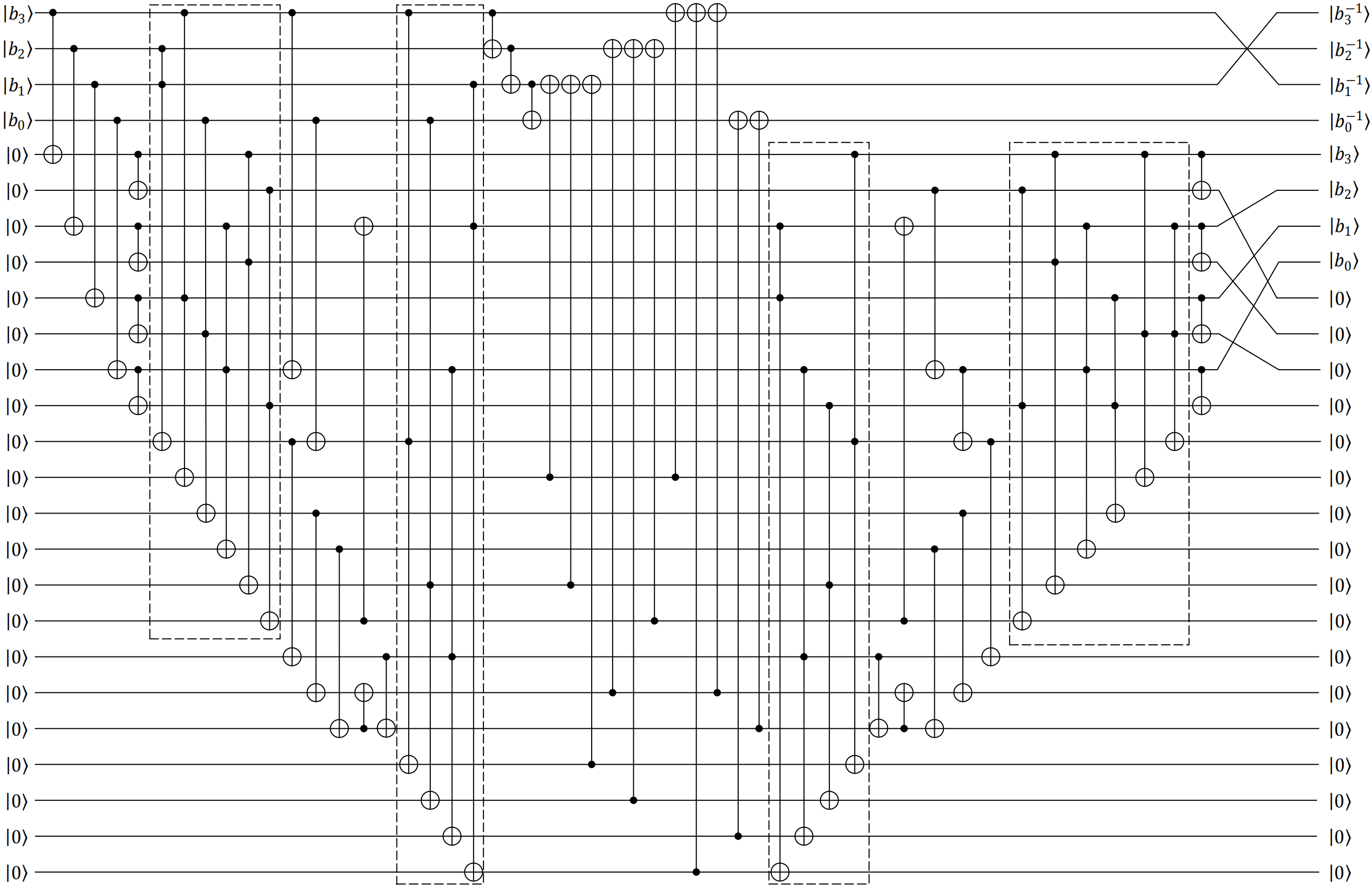}
    \caption{The quantum circuit for ${F(2^4)}_{inv1}:\ |\boldsymbol{b}\rangle|0\rangle|0\rangle \to |\boldsymbol{b^{-1}}\rangle|\boldsymbol{b}\rangle|0\rangle$, where dashed boxes indicate the $Toffoli$ gates can be applied in parallel.}
    \label{fig5-5}
\end{figure*}

Based on Eq.(\ref{ali6-1}), the quantum circuit can be redesigned in Figure \ref{fig5-5} and is denoted as ${F(2^4)}_{inv1}:\ |\boldsymbol{b}\rangle|0\rangle|0\rangle \to |\boldsymbol{b^{-1}}\rangle|\boldsymbol{b}\rangle|0\rangle$.

The quantum resource estimation for implementing the multiplicative inverse in $F(2^4)$ is summarized in Table \ref{tab2}, and the quantum resource estimation after decomposing the $Toffoli$ gates into combinations of $T$ gates and $Clifford$ gates is summarized in Table \ref{tab2-1}.

\begin{table}[!t]
\caption{Quantum resource estimation for multiplicative inversion in $F(2^4)$. $\#qubits$ denotes qubit count; $\#Toffoli$, $\#CNOT$, and $\#NOT$ are the counts of respective gates.\label{tab2}}
\centering
\footnotesize
\renewcommand{\arraystretch}{1.3}
\setlength{\tabcolsep}{3pt}
\begin{tabular}{|c|c|c|c|c|c|}
\hline
 & $\#qubits$ & $\#Toffoli$ & $\#CNOT$ & $\#NOT$ & $Toffoli$ depth \\
\hline
${F(2^4)}_{\text{inv0}}$ & 5 & 7 & 7 & 0 & 7 \\
\hline
${F(2^4)}_{\text{inv1}}$ & 25 & 20 & 42 & 0 & 4 \\
\hline
\end{tabular}
\end{table}

\begin{table}[!t]
\caption{Quantum resource estimation for multiplicative inversion in $F(2^4)$ after decomposing the $Toffoli$ gates into combinations of $T$ gates and $Clifford$ gates. $\#T$ and $\#Clifford$ are the counts of respective gates.}
\label{tab2-1}
\centering
\footnotesize
\renewcommand{\arraystretch}{1.3}
\setlength{\tabcolsep}{10pt}
\begin{tabular}{|c|c|c|c|c|}
\hline
 & $\#qubits$ & $\#T$ & $\#Clifford$ & $T$ depth \\
\hline
${F(2^4)}_{{inv0}}$ & 7 & 24 & 129 & 6 \\
\hline
${F(2^4)}_{{inv1}}$ & 29 & 40 & 222 & 2 \\
\hline
\end{tabular}
\end{table}

\subsection{Quantum circuit implementation of $q^2 \times \lambda$}
\begin{figure}[h!]
    \centering
    \includegraphics[width=0.3\textwidth]{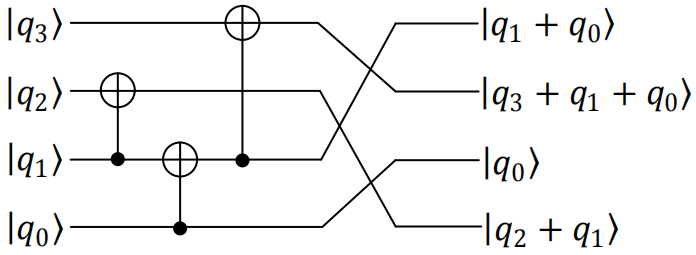}
    \caption{The quantum circuit for $U_{q^2 \times \lambda}:\ |q\rangle \to |q^2 \times \lambda \rangle$.}
    \label{fig6}
\end{figure}

 The element $q$ in $F(2^4)$ can be written as $q=q_3x^3+q_2x^2+q_1x+q_0$, where $q_i\ (i \in {0,1,2,3})$ is an element in $F(2^4)$ and $\lambda=x^3+x^2+x \in F(2^4)$ is defined in Eq.(\ref{ali2}). Through a calculation, $q^2 \times \lambda$ is expressed as
 \begin{equation}
 \label{ali6}
 q^2 \times \lambda=(q_1+q_0)x^3+(q_3+q_1+q_0)x^2+q_0x+(q_2+q_1).
\end{equation}
 Based on Eq.(\ref{ali6}), we can derive a quantum circuit for $U_{q^2 \times \lambda}:\ |q\rangle \to |q^2 \times \lambda \rangle$ in Figure \ref{fig6}. The $U_{q^2 \times \lambda}$ can be implemented with 4 qubits and 3 $CNOT$ gates.

\subsection{Quantum circuit implementations of multiplication in $F(2^4)$}
In $F(2^4)$, any element of $a$ and $b$ can be written as
\begin{equation*}
\begin{cases}
a=a_3y^3+a_2y^2+a_1y+a_0, \\
b=b_3y^3+b_2y^2+b_1y+b_0.
\end{cases}
\end{equation*}
$a \cdot b$ is expressed as
  \begin{equation}
\label{ali7}
 a \cdot b =(ab)_3y^3+(ab)_2y^2+(ab)_1y+(ab)_0,
\end{equation}
where $(ab)_i\ (i \in {0,1,2,3})$ is the $i$-th term of $a \cdot b$ and\\
$(ab)_3=(a_3+a_2+a_1+a_0)(b_3+b_2+b_1+b_0)+(a_3+a_2)(b_3+b_2)+(a_3+a_1)(b_3+b_1)+(a_2+a_0)(b_2+b_0)+(a_1+a_0)(b_1+b_0)+a_2b_2+a_1b_1+a_0b_0$,\\
$(ab)_2=(a_3+a_2)(b_3+b_2)+(a_2+a_0)(b_2+b_0)+a_1b_1+a_0b_0$,\\
$(ab)_1=(a_3+a_2)(b_3+b_2)+(a_3+a_1)(b_3+b_1)+(a_1+a_0)(b_1+b_0)+a_0b_0$,\\
$(ab)_0=(a_3+a_1)(b_3+b_1)+a_3b_3+a_2b_2+a_1b_1+a_0b_0$.

\begin{figure*}[!t]
    \centering
    \includegraphics[width=1\textwidth]{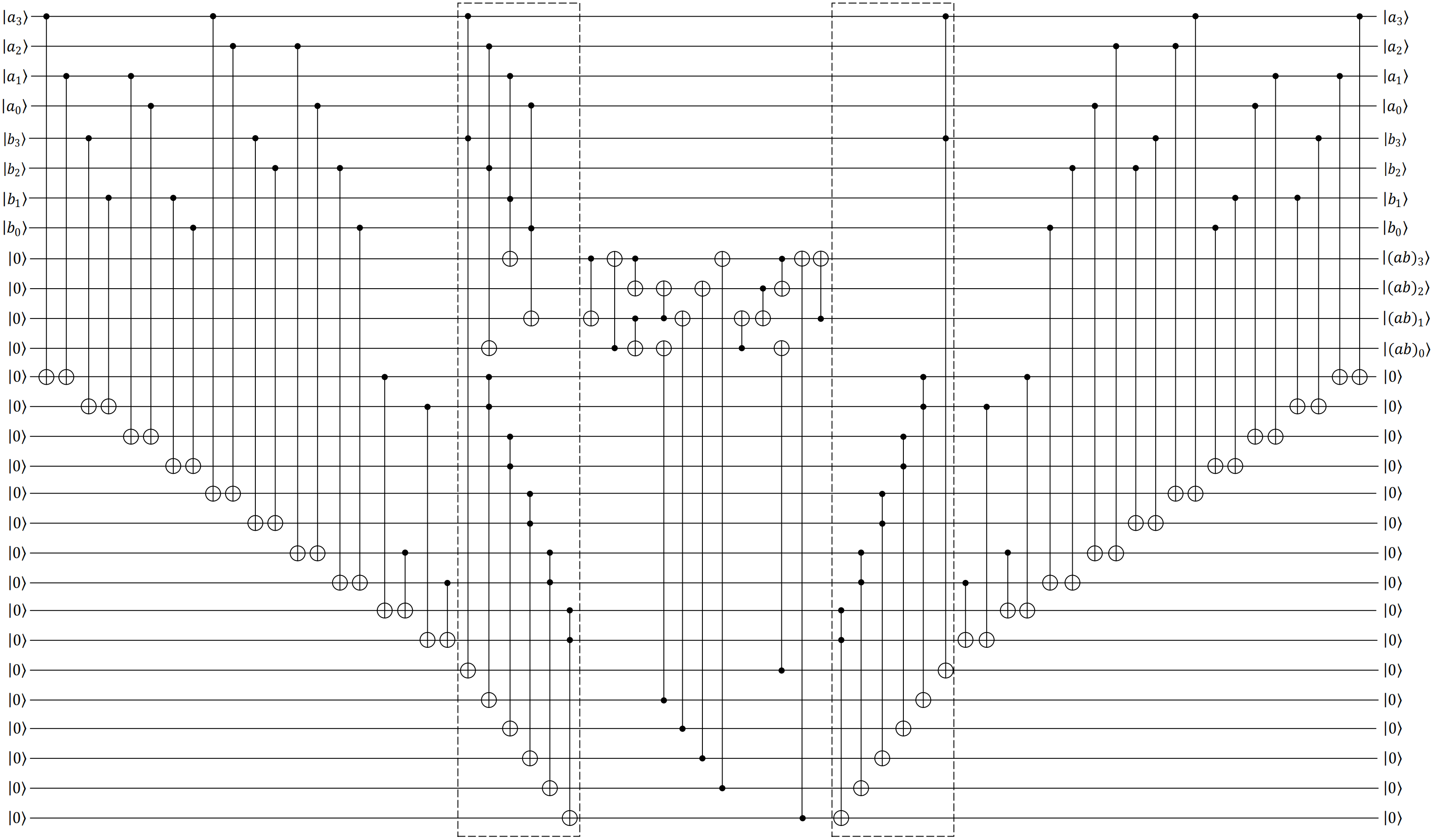}
    \caption{The quantum circuit for $Mul_0:\ |a\rangle|b\rangle|0\rangle|0\rangle \to |a\rangle|b\rangle|a \cdot b\rangle|0\rangle$.}
    \label{fig7}
\end{figure*}

Based on Eq.(\ref{ali7}), we present a quantum circuit for $Mul_0:\ |a\rangle|b\rangle|0\rangle|0\rangle \to |a\rangle|b\rangle|a \cdot b\rangle|0\rangle$ of multiplication in $F(2^4)$ in Figure \ref{fig7}. $Mul_0$ can be implemented with 28 qubits, 15 $Toffoli$ gates and 55 $CNOT$ gates. The $Toffoli$ depth of $Mul_0$ is 2. After decomposing the $Toffoli$ gates in $Mul_0$, it is found that implementing $Mul_0$ requires 36 qubits, 36 $T$ gates and 196 $Clifford$ gates. The $T$ depth of $Mul_0$ is 1. It is worth noting that $Mul_0^\dagger$ is the inverse operation of $Mul_0$. After decomposing the $Toffoli$ gates in $Mul_0^\dagger$ into $T$ and $Clifford$ gates, the implementation of $Mul_0^\dagger$ requires 36 qubits, 24 $T$ gates, and 184 $Clifford$ gates. The $T$ depth of $Mul_0^\dagger$ is 1.

\begin{figure}[h!]
    \centering
    \includegraphics[width=0.45\textwidth]{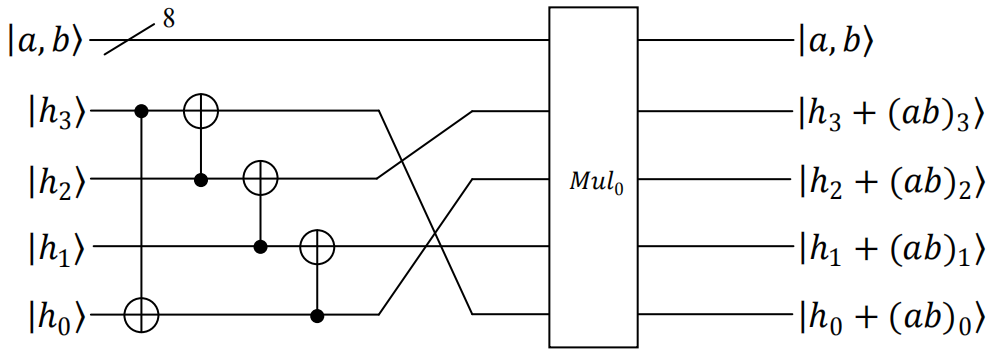}
    \caption{The quantum circuit for $Mul_1:\ |a\rangle|b\rangle|h\rangle|0\rangle \to |a\rangle|b\rangle|h+a \cdot b\rangle|0\rangle$.}
    \label{fig8}
\end{figure}

Since the key expansion process involves the SubBytes operation, the result of applying SubBytes to the current state must be XORed with another state while keeping the current state unchanged. In this case, $Mul_0$ is not applicable, so a new multiplication operation, $Mul_1:\ |a\rangle|b\rangle|h\rangle|0\rangle \to |a\rangle|b\rangle|h+a \cdot b\rangle|0\rangle$, needs to be constructed. $Mul_1$ is shown in Figure \ref{fig8}, where $Mul_0$ denotes the quantum circuit shown in Figure \ref{fig7}. $Mul_1$ can be implemented with 28 qubits, 15 $Toffoli$ gates and 59 $CNOT$ gates. The $Toffoli$ depth of $Mul_1$ is 2. After decomposing the $Toffoli$ gates in $Mul_1$, it is found that implementing $Mul_1$ requires 37 qubits, 36 $T$ gates and 200 $Clifford$ gates. The $T$ depth of $Mul_1$ is 1.

\begin{figure}[h!]
    \centering
    \includegraphics[width=0.45\textwidth]{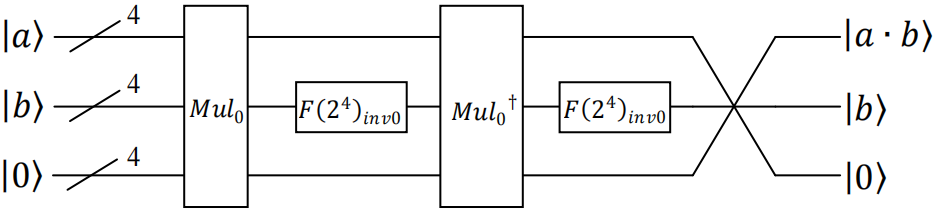}
    \caption{The quantum circuit for $Mul_2:\ |a\rangle|b\rangle|0\rangle \to |a \cdot b\rangle|b\rangle|0\rangle$.}
    \label{fig9}
\end{figure}

During the encryption process of AES, the optimal implementation of $SubBytes$ is to act directly on the qubits associated with the current state, which necessitates the design of an in-place quantum circuit for the S-box. Accordingly, the quantum circuit for multiplication in $F(2^4)$ should also be in-place. However, the previously designed multiplication circuits $Mul_0$ and $Mul_1$ perform the result on ancillary qubits, following an out-of-place design strategy. Therefore, referring to Ref.\cite{r36}, we present an in-place quantum circuit for $Mul_2: |a\rangle|b\rangle|0\rangle \to |a \cdot b\rangle|b\rangle|0\rangle$, as shown in Figure\ref{fig9}, where ${F(2^4)}_{inv0}$ denotes the quantum circuits in Section \ref{sect3.2}, and $Mul_0^\dagger$ is the inverse process of $Mul_0$.

\begin{figure}[h!]
    \centering
    \includegraphics[width=0.45\textwidth]{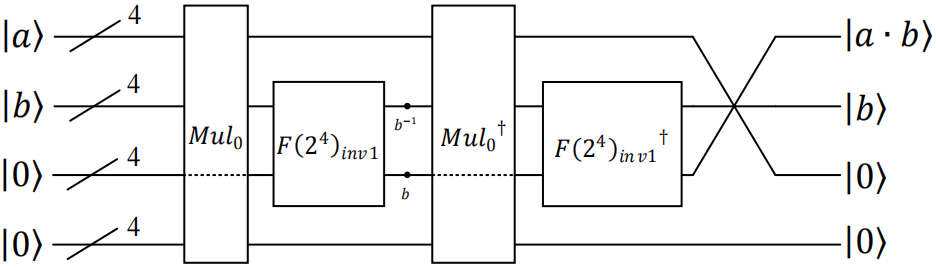}
    \caption{The new quantum circuit for $Mul_2:\ |a\rangle|b\rangle|0\rangle \to |a \cdot b\rangle|b\rangle|0\rangle$.}
    \label{fig9-1}
\end{figure}

Moreover, if five additional clean ancillary qubits can be introduced into $Mul_2$ (shown in Figure \ref{fig9}), ${F(2^4)}_{inv0}$ in $Mul_2$ can be replaced by ${F(2^4)}_{inv1}$ (shown in Figure \ref{fig9-1}), which increases the number of $Toffoli$ gates by 26 but decreases the $Toffoli$ depth by 8.

The quantum resources required for implementing $Mul_2$ are summarized in Table \ref{tab01}, and the quantum resource estimation after decomposing the $Toffoli$ gates into combinations of $T$ gates and $Clifford$ gates is summarized in Table \ref{tab01-1}.

\begin{table}[!t]
\caption{Quantum resource estimation for implementing $Mul_2$.\label{tab01}}
\centering
\footnotesize
\renewcommand{\arraystretch}{1.3}
\setlength{\tabcolsep}{3pt}
\begin{tabular}{|c|c|c|c|c|c|}
\hline
Operation & $\#qubits$ & $\#Toffoli$ & $\#CNOT$ & $\#NOT$ & $Toffoli$ depth \\
\hline
\multirow{2}{*}{$Mul_2$}
    & 28 & 44 & 124 & 0 & 18 \\
\cline{2-6}
    & 33 & 70 & 194 & 0 & 12 \\
\hline
\end{tabular}
\end{table}

\begin{table}[!t]
\caption{Quantum resource estimation for implementing $Mul_2$ after decomposing the $Toffoli$ gates into combinations of $T$ gates and $Clifford$ gates.}
\label{tab01-1}
\centering
\footnotesize
\renewcommand{\arraystretch}{1.3}
\setlength{\tabcolsep}{10pt}
\begin{tabular}{|c|c|c|c|c|}
\hline
Operation & $\#qubits$ & $\#T$ & $\#Clifford$ & $T$ depth \\
\hline
\multirow{2}{*}{$Mul_2$}
    & 36 & 108 & 638 & 14 \\
\cline{2-5}
    & 40 & 140 & 824 & 6 \\
\hline
\end{tabular}
\end{table}

\subsection{The quantum circuits of S-box}
This section presents three distinct quantum circuits for implementing the S-box, depending on varying target qubit states.

\subsubsection{The quantum circuit for $c_1:\ |a\rangle|0\rangle|0\rangle \to |a\rangle|S(a)\rangle|0\rangle$} \label{Sect3.5.1}
When the target qubit is $|0\rangle$, the quantum circuit for $c_1:\ |a\rangle|0\rangle|0\rangle \to |a\rangle|S(a)\rangle|0\rangle$ can be provided. Based on Eq.(\ref{ali3}), the S-box can be computed by realizing the quantum circuits of $M$, $(Ma)^{-1}$, $AM^{-1}$ and adding a vector $c$ in order. The quantum circuits of $M$ and $AM^{-1}$ have been presented in Section \ref{Sect3.1}. The addion of the vector $c$ can be implemented with only 4 $NOT$ gates. The remaining component is the quantum circuit that computes the multiplicative inverse $(Ma)^{-1}$ in $F((2^4)^2)$. Therefore, we first construct the quantum circuit for $U_{inv0}:\ |p\rangle|0\rangle|0\rangle \to |p\rangle|p^{-1}\rangle|0\rangle$ that implements the multiplicative inverse $(Ma)^{-1}$ in $F((2^4)^2)$, and the quantum circuit for $c_1$ is introduced subsequently.

\begin{figure}[h!]
    \centering
    \includegraphics[width=0.45\textwidth]{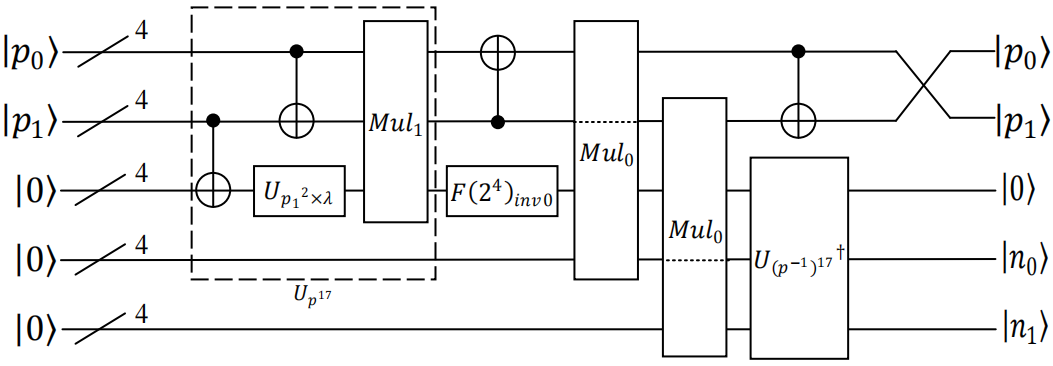}
    \caption{The quantum circuit for $U_{inv0}^1:\ |p\rangle|0\rangle|0\rangle \to |p\rangle|p^{-1}\rangle|0\rangle$, where the group of operations within the dashed box is collectively denoted as \( U_{p^{17}} \), which is used to compute \( p^{17} \) from \( p_0 \) and \( p_1 \).}
    \label{fig10}
\end{figure}

Any element in $F((2^4)^2)$ can be denoted as $p=p_0+p_1x$, where $p_0,\ p_1 \in F(2^4)$. The multiplicative inverse of \( p \) in \( F((2^4)^2) \), denoted as \( p^{-1} \), is represented as \( p^{-1} = n_0 + n_1x \) for distinction. The value of \( p^{-1} \) can be determined through a four-part process involving $(p^{17})^{-1}(p_0+p_1)$, $(p^{17})^{-1}p_1x$, ${p_1}^2 \times \lambda$, and $(p_0+p_1)p_0$. Therefore, by combining the quantum circuits of $F(2^4)_{inv0}$ (see Section \ref{sect3.2}), $Mul_0$ (see Figure \ref{fig7}), $Mul_1$ (see Figure \ref{fig8}), and $U_{q^2 \times \lambda}$ (see Figure \ref{fig6}), the quantum circuit $U_{inv0}^1:\ |p\rangle|0\rangle|0\rangle \to |p\rangle|p^{-1}\rangle|0\rangle$ for computing the multiplicative inversion in $F((2^4)^2)$ can be constructed as shown in Figure \ref{fig10}. The operation \( {U_{(p^{-1})^{17}}}^\dagger \) in Figure \ref{fig10} is the inverse of \( U_{(p^{-1})^{17}} \), where \( U_{(p^{-1})^{17}} \) computes \( (p^{-1})^{17} \) using \( n_0 \) and \( n_1 \).

\begin{figure}[h!]
    \centering
    \includegraphics[width=0.45\textwidth]{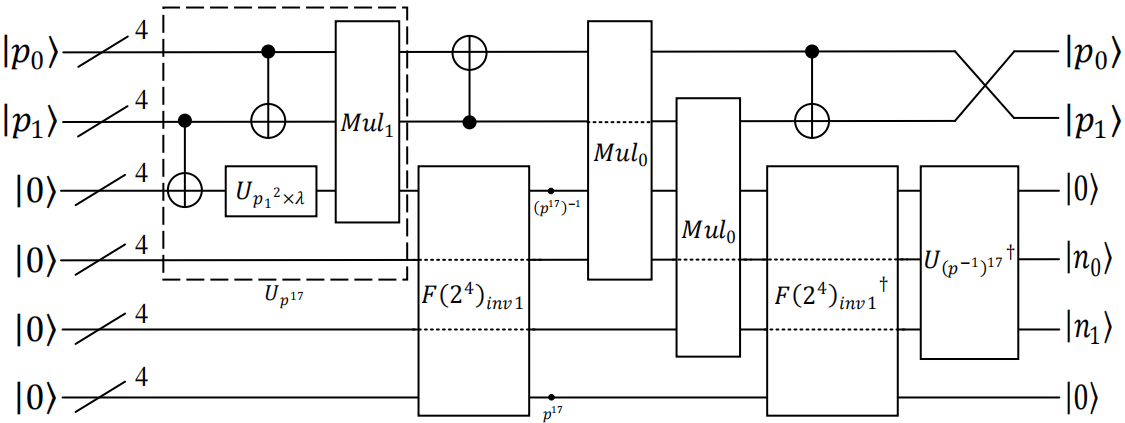}
    \caption{The quantum circuit for $U_{inv0}^2:\ |p\rangle|0\rangle|0\rangle \to |p\rangle|p^{-1}\rangle|0\rangle$.}
    \label{fig10-1}
\end{figure}

If five clean ancillary qubits are added to the $U_{inv1}^1$ (as shown in Figure \ref{fig10}), the module $F(2^4)_{inv0}$ can be replaced with $F(2^4)_{inv1}$, resulting in $U_{inv0}^2$ (as shown in Figure \ref{fig10-1}).

\begin{figure}[h!]
    \centering
    \includegraphics[width=0.35\textwidth]{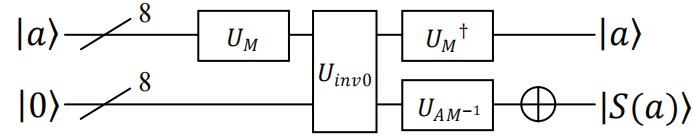}
    \caption{The quantum circuit for $c_1:\ |a\rangle|0\rangle|0\rangle \to |a\rangle|S(a)\rangle|0\rangle$.}
    \label{fig11}
\end{figure}

By combining the quantum circuit of $U_M$ (see Figure \ref{fig2}), $U_{inv0}$ ($U_{inv0}$ consists of $U_{inv0}^1$ shown in Figure \ref{fig10} and $U_{inv0}^2$ shown in Figure \ref{fig10-1}) , and $U_{AM^{-1}}$ (see Figure \ref{fig3}), the quantum circuit for $c_1:\ |a\rangle|0\rangle|0\rangle \to |a\rangle|S(a)\rangle|0\rangle$ can be constructed as shown in Figure \ref{fig11}, where ${U_M}^\dagger$ is the inverse process of $U_M$.

The quantum resources required for implementing $c_1$ are summarized in Table \ref{tab3}, where $c_1^1$ indicates that the $U_{inv0}$ used in $c_1$ is $U_{inv0}^1$, while $c_1^2$ indicates that the $U_{inv0}$ used in $c_1$ is $U_{inv0}^2$. The quantum resource estimation after decomposing the $Toffoli$ gates into combinations of $T$ gates and $Clifford$ gates is summarized in Table \ref{tab3-1}.

\begin{table}[!t]
\caption{Quantum resource estimation for implementing $c_1$.}
\label{tab3}
\centering
\footnotesize
\renewcommand{\arraystretch}{1.3}
\setlength{\tabcolsep}{3pt}
\begin{tabular}{|c|c|c|c|c|c|}
\hline
Operation & $\#qubits$ & $\#Toffoli$ & $\#CNOT$ & $\#NOT$ & $Toffoli$ depth \\
\hline
$c_1^1$ & 36 & 67 & 300 & 4 & 15 \\
\hline
$c_1^2$ & 41 & 100 & 377 & 4 & 16 \\
\hline
\end{tabular}
\end{table}

\begin{table}[!t]
\caption{Quantum resource estimation for implementing $c_1$ after decomposing the $Toffoli$ gates into combinations of $T$ gates and $Clifford$ gates.}
\label{tab3-1}
\centering
\footnotesize
\renewcommand{\arraystretch}{1.3}
\setlength{\tabcolsep}{10pt}
\begin{tabular}{|c|c|c|c|c|}
\hline
Operation & $\#qubits$ & $\#T$ & $\#Clifford$ & $T$ depth \\
\hline
$c_1^1$ & 44 & 168 & 990 & 10 \\
\hline
$c_1^2$ & 48 & 224 & 1305 & 8 \\
\hline
\end{tabular}
\end{table}

\subsubsection{The quantum circuit for $c_2:\ |a\rangle|b\rangle|0\rangle \to |a\rangle|b \oplus S(a)\rangle|0\rangle$} \label{Sect3.5.2}
When the target qubit is $|b\rangle$ instead of $|0\rangle$, the given quantum circuit can be denoted as $c_2:\ |a\rangle|b\rangle|0\rangle \to |a\rangle|b \oplus S(a)\rangle|0\rangle$. By this time, the quantum circuit $U_{inv0}$ ($U_{inv0}$ consists of $U_{inv0}^1$ shown in Figure \ref{fig10} and $U_{inv0}^2$ shown in Figure \ref{fig10-1}), which is used to compute the multiplicative inverse $(Ma)^{-1}$ in $F((2^4)^2)$, is no longer suitable for constructing $c_2$. Therefore, we need to design a new quantum circuit $U_{inv1}:\ |p\rangle|h\rangle|0\rangle \to |p\rangle|h+p^{-1}\rangle|0\rangle$ to construct $c_2$, and the quantum circuit for $c_2$ is introduced subsequently.

\begin{figure}[h!]
    \centering
    \includegraphics[width=0.45\textwidth]{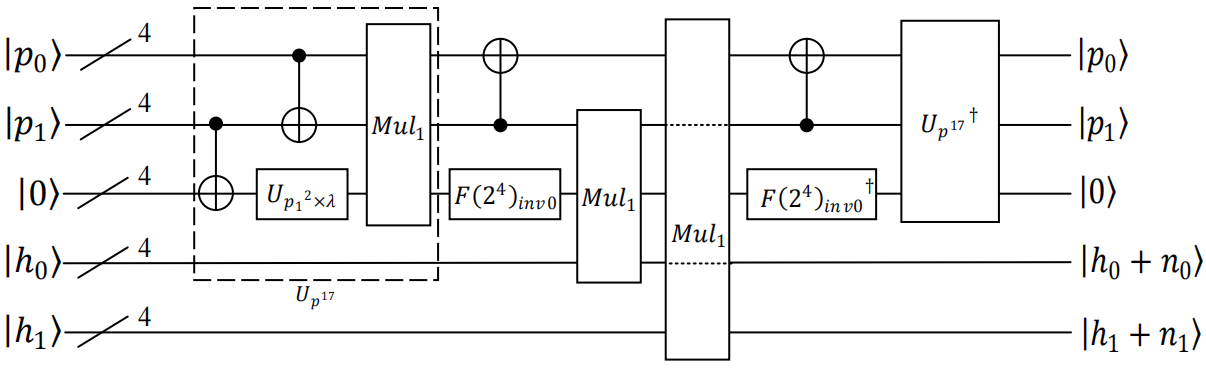}
    \caption{The quantum circuit for $U_{inv1}^1:\ |p\rangle|h\rangle|0\rangle \to |p\rangle|h+p^{-1}\rangle|0\rangle$.}
    \label{fig12}
\end{figure}

By combining the quantum circuits of $U_{q^2 \times \lambda}$ (see Figure \ref{fig6}), ${Mul_1}$ (see Figure \ref{fig8}), and ${F(2^4)_{inv0}}$ (see in Section \ref{sect3.2}), the quantum circuit $U_{inv1}^1:\ |p\rangle|h\rangle|0\rangle \to |p\rangle|h+p^{-1}\rangle|0\rangle$ for computing the multiplicative inverse in $F((2^4)^2)$ can be constructed as shown in Figure \ref{fig12}, where ${{F(2^4)}_{inv0}}^\dagger$ is the inverse process of ${F(2^4)}_{inv0}$.

\begin{figure}[h!]
    \centering
    \includegraphics[width=0.45\textwidth]{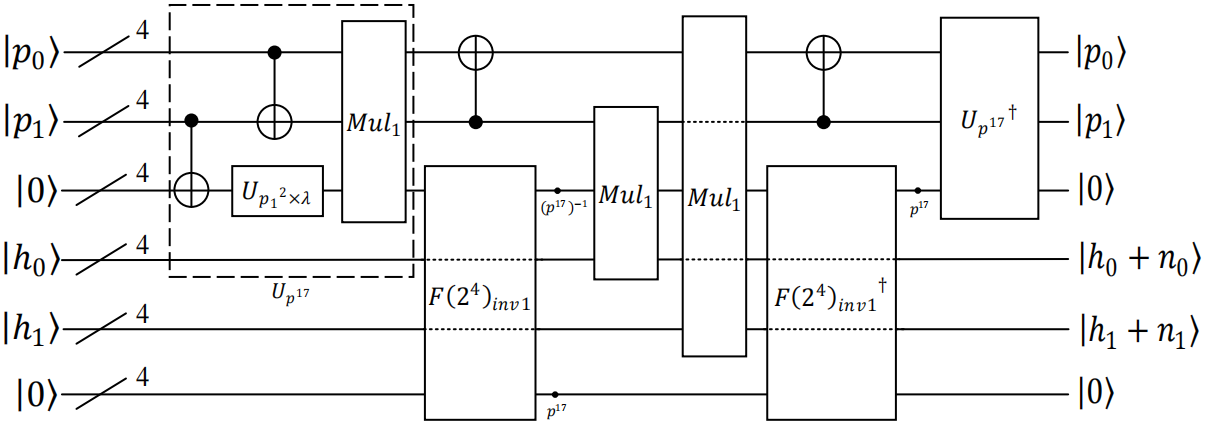}
    \caption{The quantum circuit for $U_{inv1}^2:\ |p\rangle|h\rangle|0\rangle \to |p\rangle|h+p^{-1}\rangle|0\rangle$.}
    \label{fig12-1}
\end{figure}

Similar to $U_{inv0}^1$ (as shown in Figure \ref{fig10}), if five clean ancillary qubits are added to the $U_{inv1}^1$ quantum circuit shown in Figure \ref{fig12}, the module $F(2^4)_{inv0}$ can be replaced with $F(2^4)_{inv1}$, resulting in $U_{inv1}^2$ (as shown in Figure \ref{fig12-1}). Although this replacement introduces 26 additional $Toffoli$ gates, the $Toffoli$ depth of the circuit is reduced by 8.

\begin{figure}[h!]
    \centering
    \includegraphics[width=0.45\textwidth]{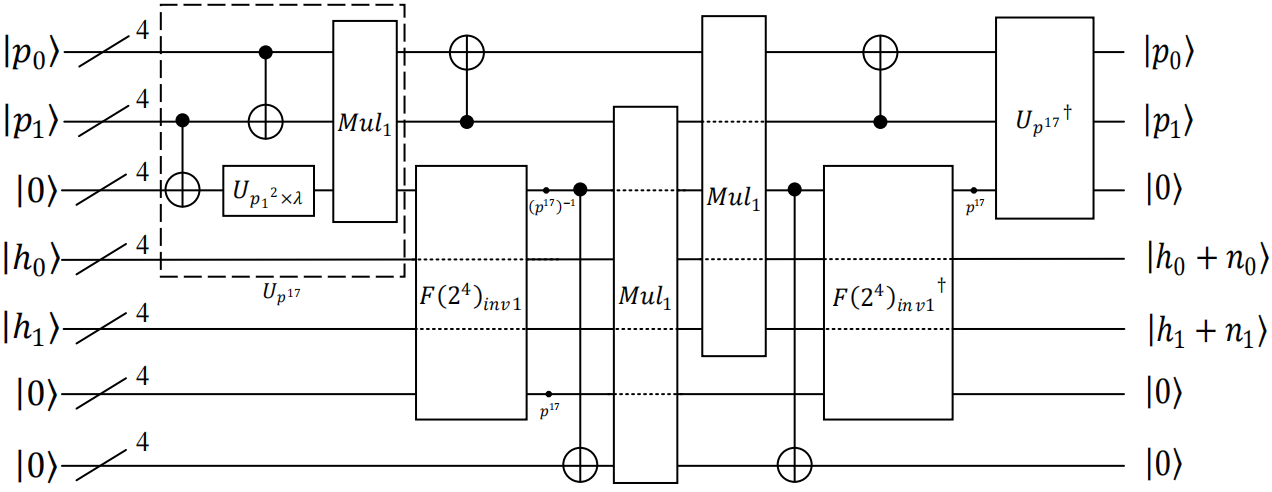}
    \caption{The quantum circuit for $U_{inv1}^3:\ |p\rangle|h\rangle|0\rangle \to |p\rangle|h+p^{-1}\rangle|0\rangle$.}
    \label{fig12-2}
\end{figure}

We observe that by adding 19 more clean ancillary qubits, the two $Mul_1$ in the $U_{inv1}^2$ shown in Figure \ref{fig12-1} can be executed in parallel, resulting in $U_{inv1}^3$ (as shown in Figure \ref{fig12-2}). Although this replacement introduces 8 additional $CNOT$ gates, the $Toffoli$ depth of the circuit is reduced by 2.

\begin{figure}[h!]
    \centering
    \includegraphics[width=0.35\textwidth]{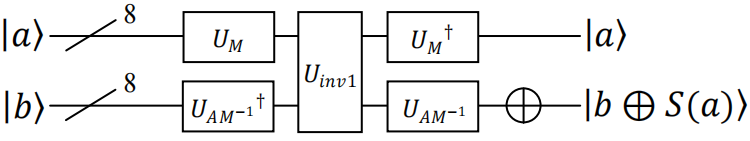}
    \caption{The quantum circuit for $c_2:\ |a\rangle|b\rangle|0\rangle \to |a\rangle|b \oplus S(a)\rangle|0\rangle$.}
    \label{fig13}
\end{figure}

By combining the quantum circuits of $U_M$ (see Figure \ref{fig2}), $U_{inv1}$ ($U_{inv1}$ consists of $U_{inv1}^1$ shown in Figure \ref{fig12}, $U_{inv1}^2$ shown in Figure \ref{fig12-1}, and $U_{inv1}^2$ shown in Figure \ref{fig12-2}), and $U_{AM^{-1}}$ (see Figure \ref{fig3}), the quantum circuit for $c_2:\ |a\rangle|b\rangle|0\rangle \to |a\rangle|b \oplus S(a)\rangle|0\rangle$ can be constructed as shown in Figure \ref{fig13}, where ${U_M}^\dagger$ is the inverse process of $U_M$ and ${U_{AM^{-1}}}^\dagger$ is the inverse process of $U_{AM^{-1}}$.

The quantum resources required for implementing $c_2$ are shown in Table \ref{tab4}, where $c_2^1$ indicates that the $U_{inv1}$ used in $c_2$ is $U_{inv1}^1$, $c_2^2$ indicates that the $U_{inv1}$ used in $c_2$ is $U_{inv1}^2$, and $c_2^3$ indicates that the $U_{inv1}$ used in $c_2$ is $U_{inv1}^3$. The quantum resource estimation after decomposing the $Toffoli$ gates into combinations of $T$ gates and $Clifford$ gates is summarized in Table \ref{tab4-1}.

\begin{table}[!t]
\caption{Quantum resource estimation for implementing $c_2$.}
\label{tab4}
\centering
\footnotesize
\renewcommand{\arraystretch}{1.3}
\setlength{\tabcolsep}{3pt}
\begin{tabular}{|c|c|c|c|c|c|}
\hline
Operation & $\#qubits$ & $\#Toffoli$ & $\#CNOT$ & $\#NOT$ & $Toffoli$ depth \\
\hline
$c_2^1$ & 36 & 74 & 330 & 4 & 22 \\
\hline
$c_2^2$ & 41 & 100 & 400 & 4 & 16 \\
\hline
$c_2^3$ & 60 & 100 & 408 & 4 & 14 \\
\hline
\end{tabular}
\end{table}

\begin{table}[!t]
\caption{Quantum resource estimation for implementing $c_2$ after decomposing the $Toffoli$ gates into combinations of $T$ gates and $Clifford$ gates.}
\label{tab4-1}
\centering
\footnotesize
\renewcommand{\arraystretch}{1.3}
\setlength{\tabcolsep}{10pt}
\begin{tabular}{|c|c|c|c|c|}
\hline
Operation & $\#qubits$ & $\#T$ & $\#Clifford$ & $T$ depth \\
\hline
$c_2^1$ & 45 & 192 & 1142 & 16 \\
\hline
$c_2^2$ & 49 & 224 & 1328 & 8 \\
\hline
$c_2^3$ & 78 & 224 & 1336 & 7 \\
\hline
\end{tabular}
\end{table}

\subsubsection{The quantum circuit for $c_3:\ |a\rangle|0\rangle \to |S(a)\rangle|0\rangle$} \label{Sect3.5.3}
In order to reduce the number of qubits used in the quantum circuit of the S-box, this section proposes a quantum circuit implementation for $c_3:\ |a\rangle|0\rangle \to |S(a)\rangle|0\rangle$. Now, the quantum circuits for $U_{inv0}$ ($U_{inv0}$ consists of $U_{inv0}^1$ shown in Figure \ref{fig10} and $U_{inv0}^2$ shown in Figure \ref{fig10-1}) and $U_{inv1}$ ($U_{inv1}$ consists of $U_{inv1}^1$ shown in Figure \ref{fig12}, $U_{inv1}^2$ shown in Figure \ref{fig12-1}, and $U_{inv1}^2$ shown in Figure \ref{fig12-2}), are no longer suitable for constructing $c_3$. Therefore, we need to design a new quantum circuit $U_{inv2}:\ |p\rangle|0\rangle \to |p^{-1}\rangle|0\rangle$ to construct $c_3$, and the quantum circuit for $c_3$ is introduced subsequently.

\begin{figure}[h!]
    \centering
    \includegraphics[width=0.45\textwidth]{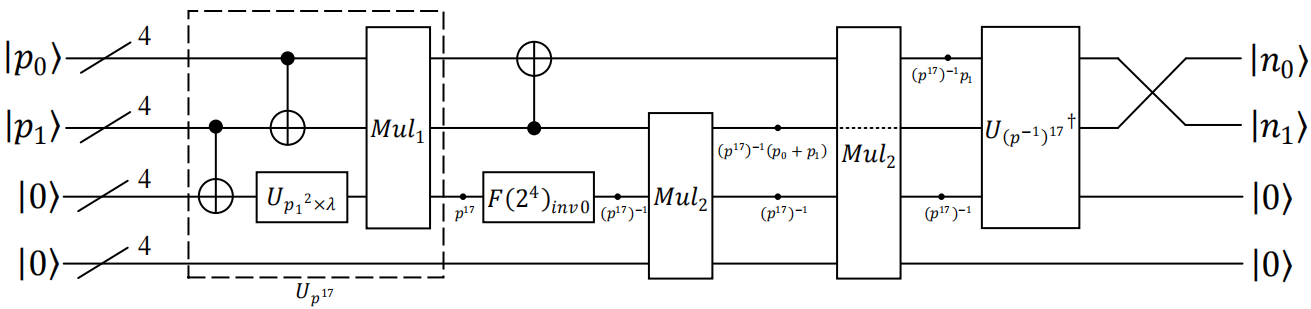}
    \caption{The quantum circuit for $U_{inv2}^1:\ |p\rangle|0\rangle \to |p^{-1}\rangle|0\rangle$, where $p^{-1}=n_0+n_1x$.}
    \label{fig14}
\end{figure}

By combining the quantum circuits of $U_{q^2 \times \lambda}$ (see Figure \ref{fig6}), $Mul_1$ (see Figure \ref{fig8}) and $Mul_2$ (see Figure \ref{fig9}), and ${F(2^4)}_{inv0}$ (see in Section \ref{sect3.2}), the quantum circuit $U_{inv2}^1:\ |p\rangle|0\rangle \to |p^{-1}\rangle|0\rangle$ for computing the multiplicative inverse in $F((2^4)^2)$ can be constructed as shown in Figure \ref{fig14}.

\begin{figure}[h!]
    \centering
    \includegraphics[width=0.45\textwidth]{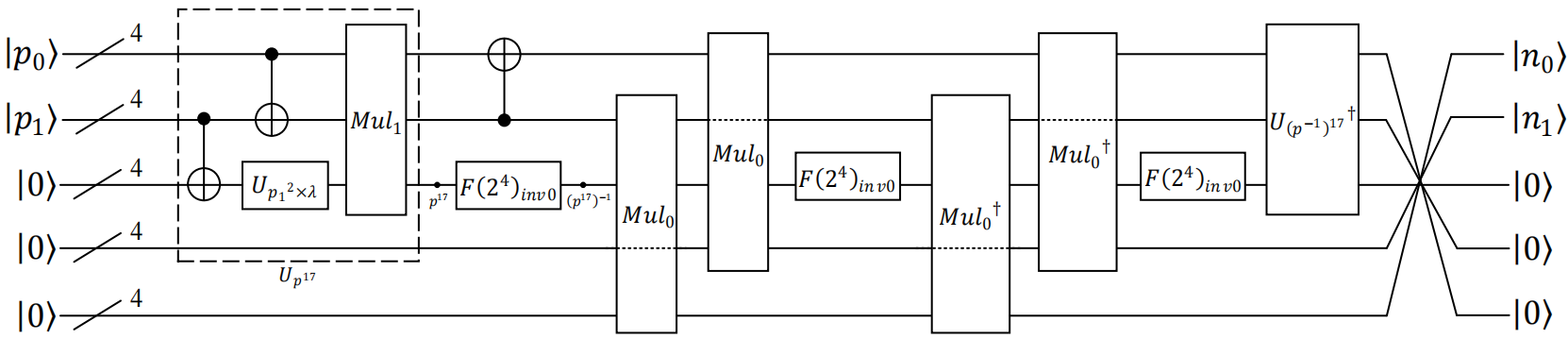}
    \caption{The quantum circuit for $U_{inv2}^2:\ |p\rangle|0\rangle \to |p^{-1}\rangle|0\rangle$.}
    \label{fig14-0}
\end{figure}

By examining the quantum circuits of $Mul_2$ (see Figure \ref{fig9}) and $U_{inv2}^1$ (see Figure \ref{fig14}), we observe that by adding four clean ancillary qubits, the two $Mul_2$ in $U_{inv2}^1$ can be reorganized, resulting in $U_{inv2}^2$ shown in Figure \ref{fig14-0}. This reorganization reduces the number of $Toffoli$ gates and $Toffoli$ depth by 14.

\begin{figure}[h!]
    \centering
    \includegraphics[width=0.45\textwidth]{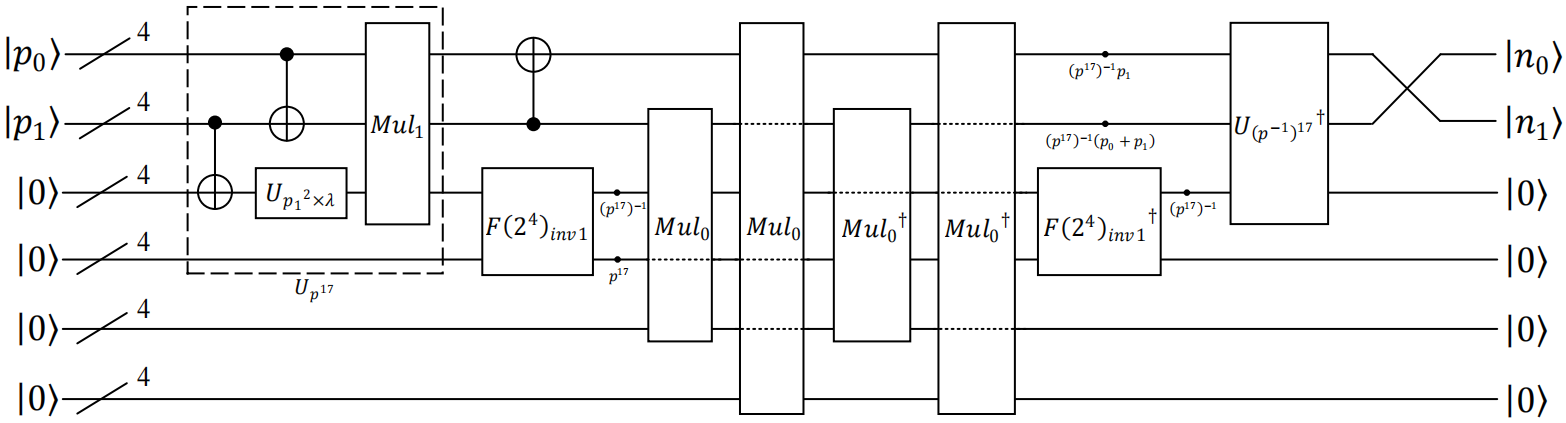}
    \caption{The quantum circuit for $U_{inv2}^3:\ |p\rangle|0\rangle \to |p^{-1}\rangle|0\rangle$.}
    \label{fig14-1}
\end{figure}

Similar to $U_{inv0}$ ($U_{inv0}$ consists of $U_{inv0}^1$ shown in Figure \ref{fig10} and $U_{inv0}^2$ shown in Figure \ref{fig10-1}) and $U_{inv1}$ ($U_{inv1}$ consists of $U_{inv1}^1$ shown in Figure \ref{fig12}, $U_{inv1}^2$ shown in Figure \ref{fig12-1}, and $U_{inv1}^2$ shown in Figure \ref{fig12-2}), if four clean ancillary qubits are added to the $U_{inv2}^2$ (shown in Figure \ref{fig14-0}), the two $F(2^4)_{inv0}$ in $U_{inv2}^2$ can be replaced with $F(2^4)_{inv1}$ (shown in Figure \ref{fig5-5}), resulting in $U_{inv2}^3$ as shown in Figure \ref{fig14-1}. Although this replacement introduces 19 additional $Toffoli$ gates, the $Toffoli$ depth of the circuit is reduced by 15.

\begin{figure}[h!]
    \centering
    \includegraphics[width=0.45\textwidth]{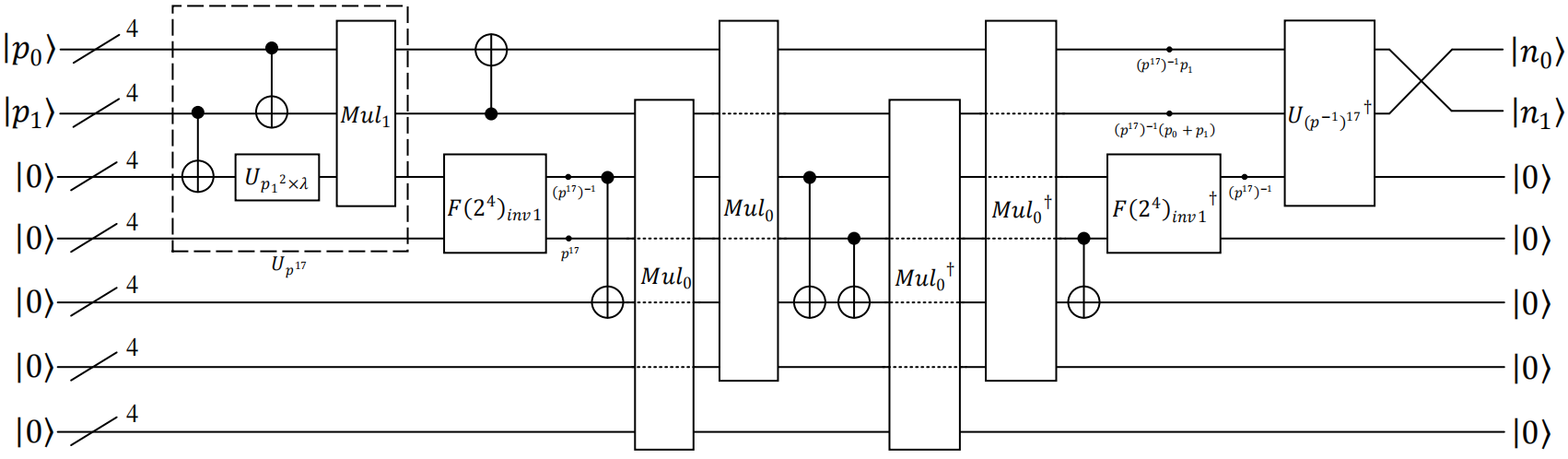}
    \caption{The quantum circuit for $U_{inv2}^4:\ |p\rangle|0\rangle \to |p^{-1}\rangle|0\rangle$.}
    \label{fig14-2}
\end{figure}

We observe that by adding 20 more clean ancillary qubits, the two $Mul_1$ in the $U_{inv1}$ (shown in Figure \ref{fig14-1}) can be executed in parallel, resulting in $U_{inv2}^4$ shown in Figure \ref{fig14-2}. Although this replacement introduces 16 additional $CNOT$ gates, the $Toffoli$ depth of the circuit is reduced by 4.

\begin{figure}[h!]
    \centering
    \includegraphics[width=0.35\textwidth]{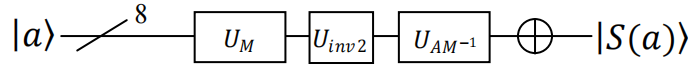}
    \caption{The quantum circuit for $c_3:\ |a\rangle|0\rangle \to |S(a)\rangle|0\rangle$.}
    \label{fig15}
\end{figure}

By combining the quantum circuits of $U_M$ (see Figure \ref{fig2}), $U_{inv2}$ ($U_{inv2}$ consists of $U_{inv2}^1$ shown in Figure \ref{fig14}, $U_{inv2}^2$ shown in Figure \ref{fig14-0}, $U_{inv2}^3$ shown in Figure \ref{fig14-1}, and $U_{inv2}^4$ shown in Figure \ref{fig14-2}) and $U_{AM^{-1}}$ (see Figure \ref{fig3}), the quantum circuit for $c_3:\ |a\rangle|0\rangle \to |S(a)\rangle|0\rangle$ can be constructed as shown in Figure \ref{fig15}.

The quantum resources required to implement $c_3$ are listed in Table \ref{tab5}, where $c_3^1$ indicates that the $U_{inv2}$ used in $c_3$ is $U_{inv2}^1$, $c_3^2$ indicates that the $U_{inv2}$ used in $c_3$ is $U_{inv2}^2$, $c_3^3$ indicates that the $U_{inv2}$ used in $c_3$ is $U_{inv2}^3$, and $c_3^4$ indicates that the $U_{inv2}$ used in $c_3$ is $U_{inv2}^4$. The quantum resource estimation after decomposing the $Toffoli$ gates into combinations of $T$ gates and $Clifford$ gates is summarized in Table \ref{tab5-1}.

\begin{table}[!t]
\caption{Quantum resource estimation for implementing $c_3$.}
\label{tab5}
\centering
\footnotesize
\renewcommand{\arraystretch}{1.3}
\setlength{\tabcolsep}{3pt}
\begin{tabular}{|c|c|c|c|c|c|}
\hline
Operation & $\#qubits$ & $\#Toffoli$ & $\#CNOT$ & $\#NOT$ & $Toffoli$ depth \\
\hline
$c_3^1$ & 32 & 125 & 424 & 4 & 47 \\
\hline
$c_3^2$ & 36 & 111 & 410 & 4 & 33 \\
\hline
$c_3^3$ & 40 & 130 & 473 & 4 & 20 \\
\hline
$c_3^4$ & 60 & 130 & 489 & 4 & 16 \\
\hline
\end{tabular}
\end{table}

\begin{table}[!t]
\caption{Quantum resource estimation for implementing $c_3$ after decomposing the $Toffoli$ gates into combinations of $T$ gates and $Clifford$ gates.}
\label{tab5-1}
\centering
\footnotesize
\renewcommand{\arraystretch}{1.3}
\setlength{\tabcolsep}{10pt}
\begin{tabular}{|c|c|c|c|c|}
\hline
Operation & $\#qubits$ & $\#T$ & $\#Clifford$ & $T$ depth \\
\hline
$c_3^1$ & 40 & 312 & 1860 & 36 \\
\hline
$c_3^2$ & 44 & 264 & 1602 & 24 \\
\hline
$c_3^3$ & 48 & 272 & 1659 & 10 \\
\hline
$c_3^4$ & 76 & 272 & 1675 & 8 \\
\hline
\end{tabular}
\end{table}

\section{The quantum circuit for AES} \label{sect4}
In this section, we will utilize the designed quantum circuits for the S-box, along with the linear structures and key expansion process, to design the quantum circuit for AES.

\subsection{The linear transformation structures of AES} \label{sect4.1}
The linear transformation structures of AES, ShiftRows, MixColumns and AddRoundKey, can all be implemented with $CNOT$ gates only. The quantum resources required for these three linear transformations are detailed below.

1. AddRoundKey: Each AddRoundKey operation involves $XOR$ing a 128-bit subkey with the current state. This can be implemented in parallel using 128 $CNOT$ gates.

2. ShiftRows: The ShiftRows operation only rearranges the output order of 16 bytes in the current state. It does not require any quantum operation for implementation.

3. MixColumns: The MixColumns operation processes 32 bits in the current state at a time and can be implemented using a $32 \times 32$ matrix. Xiang et al. \cite{r40} provides a quantum circuit for MixColumns operation with 92 $CNOT$ gates. As MixColumns can be applied to all 4 columns of the state simultaneously, and the total requirement for $CNOT$ gates is $92 \times 4 = 368$.

\subsection{The key expansion process of AES} \label{sect4.2}
\begin{figure}[h!]
    \centering
    \includegraphics[width=0.45\textwidth]{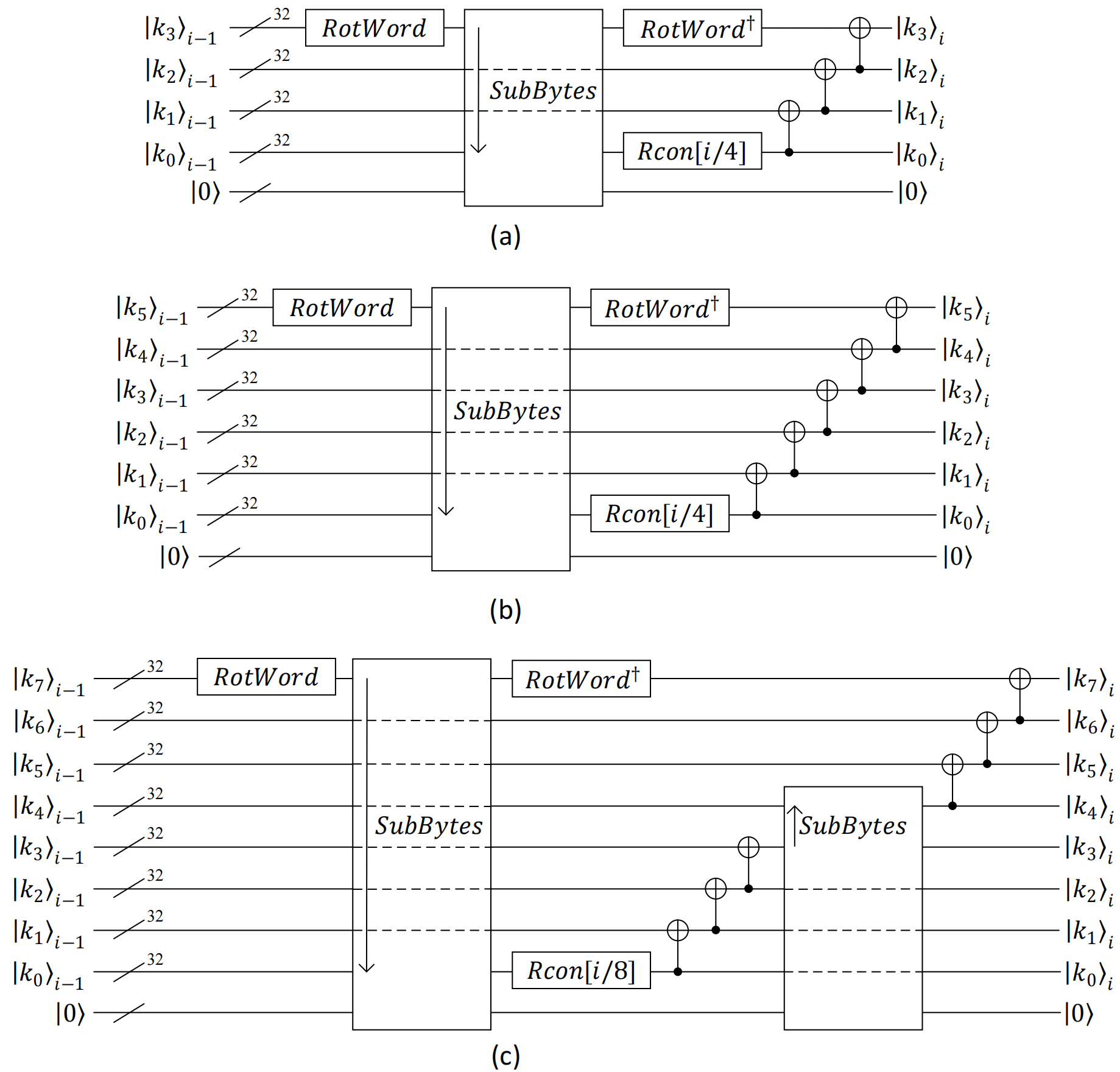}
    \caption{Three fully structures of the key expansion process \cite{r42}. (a) A full structure of the key expansion process that can be used for AES-128. (b) A full structure of the key expansion process that can be used for AES-192. (c) A full structure of the key expansion process that can be used for AES-256.}
    \label{fig16}
\end{figure}

For the key expansion process, we implement the key expansion process of AES with Jaques et al.＊s method \cite{r42}, but we use our S-box circuit for $c_2$ (see Figure \ref{fig13}). The quantum circuit as shown in Figure \ref{fig16} represents the full structure of the key expansion process, denoted as $\mathcal{KE}:\ {|k\rangle}_{i-1}|0\rangle \to |k\rangle_i|0\rangle$. The $SubBytes$ with an arrow in Figure \ref{fig16} indicates that the state at the tail of the arrow first undergoes the $SubBytes$ operation, and then the resulting value is then $XOR$ing with the state at the head of the arrow. The final result is assigned to the head of the arrow.

The initial key of AES undergoes $N_R$ rounds of the key expansion process, resulting in $(N_R+1)$ round keys. Based on Algorithms \ref{Alg1} and \ref{Alg2}, each round of key expansion process contains $RotWord$, $SubBytes$, $Rcon$ and $XOR$ four operations. The $RotWord$ can be implemented by simply rearranging the output order of the qubits, without requiring any quantum operation. The $Rcon$ is a constant (see Section \ref{Sect2.1.2}) and can be efficiently implemented with at most four $NOT$ gates for AES-128, at most five $NOT$ gates for AES-192, while at most one $NOT$ gate for AES-256 in each round of the key expansion process. The $XOR$ operation between two 32-bit words during key expansion can be implemented with 32 $CNOT$ gates. Since each round of the key expansion process operates on 128 qubits, only the relevant subset of the full structure of the key expansion process (see Figure \ref{fig16}) is needed, with other operations being neglected. Let ${\mathcal{KE}}^l_j$ represents the operation from ${|k_j\rangle}_{i-1}$ to ${|k_l\rangle}_{i-1}$ in $\mathcal{KE}:\ {|k\rangle}_{i-1}|0\rangle \to |k\rangle_i|0\rangle$, while ignoring other parts, where $0 \le j \le l$. The quantum resources required to implement the key expansion process of AES with $T$ gates and $Clifford$ gates are shown in Table \ref{tab6-1}.

\begin{table}[!t]
\caption{Quantum resources required to implement the key expansion processes of AES.}
\label{tab6-1}
\centering
\scriptsize
\renewcommand{\arraystretch}{1.3}
\setlength{\tabcolsep}{7pt}
\begin{tabular}{|c|c|c|c|c|c|c|}
\hline
 & Operation & $\#qubits$ & $\#T$ & $\#Clifford$ & $T$ depth \\
\hline
\multirow{3}{*}{AES-128}
    & $c_2^1$ & 244 & 7680 & 46659 & 160 \\
    \cline{2-6}
    & $c_2^2$ & 260 & 8960 & 54099 & 80 \\
    \cline{2-6}
    & $c_2^3$ & 376 & 8960 & 54419 & 70 \\
\hline
\multirow{3}{*}{AES-192}
    & $c_2^1$ & 308 & 8448 & 51395 & 176 \\
    \cline{2-6}
    & $c_2^2$ & 324 & 9856 & 59579 & 88 \\
    \cline{2-6}
    & $c_2^3$ & 440 & 9856 & 59931 & 77 \\
\hline
\multirow{3}{*}{AES-256}
    & $c_2^1$ & 372 & 9984 & 60639 & 208 \\
    \cline{2-6}
    & $c_2^2$ & 388 & 11648 & 70311 & 104 \\
    \cline{2-6}
    & $c_2^3$ & 504 & 11648 & 70727 & 91 \\
\hline
\end{tabular}
\end{table}

\subsection{The SubBytes operation of AES} \label{sect4.3}
The number of encryption rounds $N_R$ for AES-128, AES-192, and AES-256 is 10, 12, and 14, respectively. Since the plaintext length is 128 bits, SubBytes operation of each round function requires 16 executions of the $8 \times 8$ S-box. Since this paper focuses on implementing the AES with as few qubits as possible, the SubBytes operation during encryption is implemented using $c_3:\ |a\rangle|0\rangle \to |S(a)\rangle|0\rangle$. The quantum resources required to implement the quantum circuit for SubBytes in AES with $T$ gates and $Clifford$ gates are summarized in Table \ref{tab06-1}.

\begin{table}[!t]
\caption{Quantum resources required to implement the SubBytes operation of AES.}
\label{tab06-1}
\centering
\scriptsize
\renewcommand{\arraystretch}{1.3}
\setlength{\tabcolsep}{7pt}
\begin{tabular}{|c|c|c|c|c|c|}
\hline
 & Operation & $\#qubits$ & $\#T$ & $\#Clifford$ & $T$ depth \\
\hline
\multirow{4}{*}{AES-128}
    & $c_3^1$ & 640 & 49920 & 297600 & 360 \\
    \cline{2-6}
    & $c_3^2$ & 704 & 42220 & 256320 & 240 \\
    \cline{2-6}
    & $c_3^3$ & 768 & 43520 & 265440 & 100 \\
    \cline{2-6}
    & $c_3^4$ & 1216 & 43520 & 268000 & 80 \\
\hline
\multirow{4}{*}{AES-192}
    & $c_3^1$ & 704 & 59904 & 357120 & 432 \\
    \cline{2-6}
    & $c_3^2$ & 768 & 50688 & 307584 & 288 \\
    \cline{2-6}
    & $c_3^3$ & 832 & 52224 & 318528 & 120 \\
    \cline{2-6}
    & $c_3^4$ & 1280 & 52224 & 321600 & 96 \\
\hline
\multirow{4}{*}{AES-256}
    & $c_3^1$ & 768 & 69888 & 416640 & 504 \\
    \cline{2-6}
    & $c_3^2$ & 832 & 59136 & 358848 & 336 \\
    \cline{2-6}
    & $c_3^3$ & 869 & 60928 & 371616 & 140 \\
    \cline{2-6}
    & $c_3^4$ & 1344 & 60928 & 375200 & 112 \\
\hline
\end{tabular}
\end{table}

\subsection{Integrated quantum circuit for AES}

\begin{figure*}[!t]
    \centering
    \includegraphics[width=1.0\textwidth]{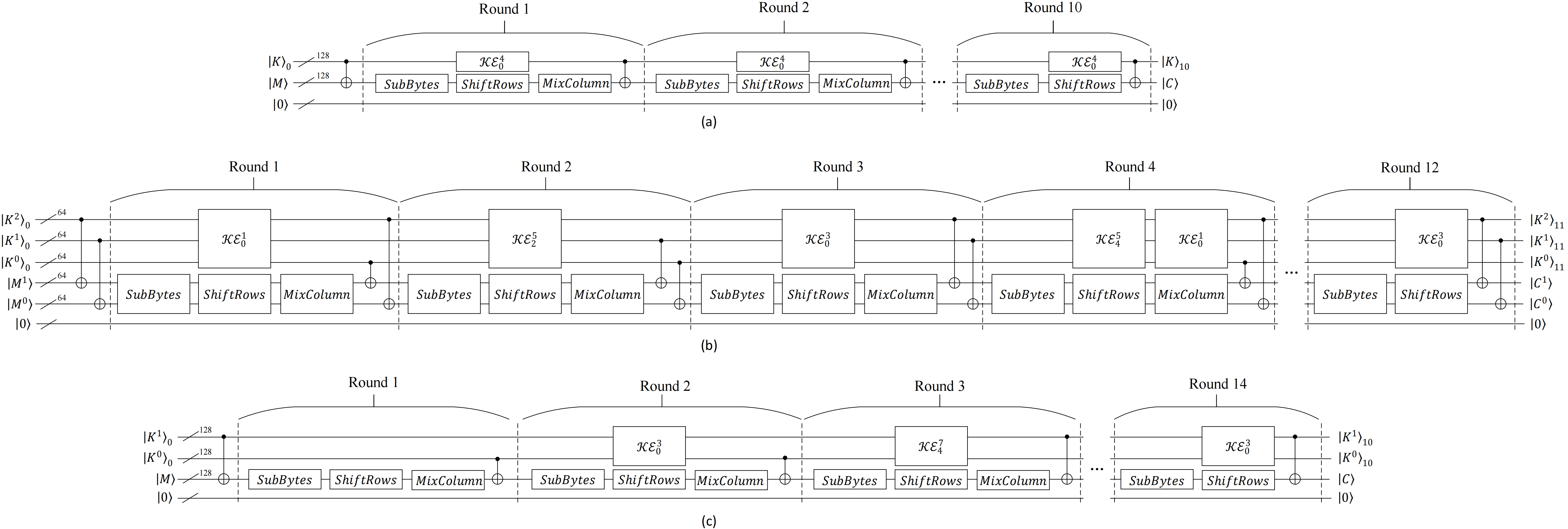}
    \caption{The encryption process of AES. (a) The encryption process of AES-128. (b) The encryption process of AES-192. (c) The encryption process of AES-256.}
    \label{fig17}
\end{figure*}

The AES consists of the encryption process and the key expansion process. The encryption process includes an initial AddRoundKey operation followed by \( N_R \) rounds of round functions, and each round function consists of SubBytes, ShiftRows, MixColumns, and AddRoundKey. The quantum circuit of the AES is presented in Figure \ref{fig17}, where $SubBytes$, $ShiftRows$ and $MixColumn$ are the operations of SubBytes, ShiftRows and MixColumns, and $\mathcal{KE}^l_j$ represents the operation from ${|k_j\rangle}_{i-1}$ to ${|k_l\rangle}_{i-1}$ in $\mathcal{KE}:\ {|k\rangle}_{i-1}|0\rangle \to |k\rangle_i|0\rangle$, while disregarding other words.

The quantum resources required to implement the SubBytes operation in the encryption process are described in Section \ref{sect4.3}, while those for ShiftRows, MixColumns and AddRoundKey are provided in Section \ref{sect4.1}. The quantum resources required for the key expansion process are detailed in Section \ref{sect4.2}. To sum up, the quantum resources required to implement the quantum circuit for AES-128, AES-192, and AES-256 with $T$ gates and $Clifford$ gates are summarized in Table \ref{tab7-1}, Table \ref{tab8-1}, and Table \ref{tab9-1}, respectively.

\begin{table}[!t]
\caption{The quantum resources required to implement the quantum circuits for AES-128, where DW($T$) is defined as the product of the $T$ depth and the circuit width (i.e., the number of qubits), and $c_2^i\text{-}c_3^j$ indicates that $c_2^i$ is used in the key expansion and $c_3^j$ is used in the SubBytes ($1 \leq i \leq 3$ and $1 \leq j \leq 4$).}
\label{tab7-1}
\centering
\scriptsize
\renewcommand{\arraystretch}{1.3}
\setlength{\tabcolsep}{7pt}
\begin{tabular}{|c|c|c|c|c|c|}
\hline
Operation & \#qubits & \#$T$ & \#$Clifford$ & $T$ depth & DW$(T)$ \\
\hline
$c_2^1$-$c_3^1$ & 884 & 57600 & 349219 & 360 & 318240 \\
\hline
$c_2^2$-$c_3^1$ & 900 & 58880 & 356659 & 360 & 324000 \\
\hline
$c_2^3$-$c_3^1$ & 1016 & 58880 & 356979 & 360 & 365760 \\
\hline
$c_2^1$-$c_3^2$ & 948 & 49920 & 307939 & 240 & 227520 \\
\hline
$c_2^2$-$c_3^2$ & 964 & 51200 & 315379 & 240 & 231360 \\
\hline
$c_2^3$-$c_3^2$ & 1080 & 51200 & 315699 & 240 & 259200 \\
\hline
$c_2^1$-$c_3^3$ & 1012 & 51200 & 317059 & 160 & 161920 \\
\hline
$c_2^2$-$c_3^3$ & 1028 & 52480 & 324499 & 100 & 102800 \\
\hline
$c_2^3$-$c_3^3$ & 1144 & 52480 & 324819 & 100 & 114400 \\
\hline
$c_2^1$-$c_3^4$ & 1460 & 51200 & 319619 & 160 & 233600 \\
\hline
$c_2^2$-$c_3^4$ & 1476 & 52480 & 327059 & 80 & 118080 \\
\hline
$c_2^3$-$c_3^4$ & 1592 & 52480 & 327379 & 80 & 127360 \\
\hline
\end{tabular}
\end{table}

\begin{table}[!t]
\caption{The quantum resources required to implement the quantum circuits for AES-192.}
\label{tab8-1}
\centering
\scriptsize
\renewcommand{\arraystretch}{1.3}
\setlength{\tabcolsep}{7pt}
\begin{tabular}{|c|c|c|c|c|c|}
\hline
Operation & \#qubits & \#$T$ & \#$Clifford$ & $T$ depth & DW$(T)$ \\
\hline
$c_2^1$-$c_3^1$ & 1012 & 68352 & 414467 & 432 & 437184 \\
\hline
$c_2^2$-$c_3^1$ & 1028 & 69760 & 422651 & 432 & 444096 \\
\hline
$c_2^3$-$c_3^1$ & 1144 & 69760 & 423003 & 432 & 494208 \\
\hline
$c_2^1$-$c_3^2$ & 1076 & 59136 & 364931 & 288 & 309888 \\
\hline
$c_2^2$-$c_3^2$ & 1092 & 60544 & 373115 & 288 & 314496 \\
\hline
$c_2^3$-$c_3^2$ & 1208 & 60544 & 373467 & 288 & 347904 \\
\hline
$c_2^1$-$c_3^3$ & 1140 & 60672 & 375875 & 176 & 200640 \\
\hline
$c_2^2$-$c_3^3$ & 1156 & 62080 & 384059 & 120 & 138720 \\
\hline
$c_2^3$-$c_3^3$ & 1272 & 62080 & 384411 & 120 & 152640 \\
\hline
$c_2^1$-$c_3^4$ & 1588 & 60672 & 378947 & 176 & 279488 \\
\hline
$c_2^2$-$c_3^4$ & 1604 & 62080 & 387131 & 96 & 153984 \\
\hline
$c_2^3$-$c_3^4$ & 1720 & 62080 & 387483 & 96 & 165120 \\
\hline
\end{tabular}
\end{table}

\begin{table}[!t]
\caption{The quantum resources required to implement the quantum circuits for AES-256.}
\label{tab9-1}
\centering
\scriptsize
\renewcommand{\arraystretch}{1.3}
\setlength{\tabcolsep}{7pt}
\begin{tabular}{|c|c|c|c|c|c|}
\hline
Operation & \#qubits & \#$T$ & \#$Clifford$ & $T$ depth & DW$(T)$ \\
\hline
$c_2^1$-$c_3^1$ & 1140 & 79872 & 484223 & 504 & 574560 \\
\hline
$c_2^2$-$c_3^1$ & 1156 & 81536 & 493895 & 504 & 582624 \\
\hline
$c_2^3$-$c_3^1$ & 1272 & 81536 & 494311 & 504 & 641088 \\
\hline
$c_2^1$-$c_3^2$ & 1204 & 69120 & 426431 & 336 & 404544 \\
\hline
$c_2^2$-$c_3^2$ & 1220 & 70784 & 436103 & 336 & 409920 \\
\hline
$c_2^3$-$c_3^2$ & 1336 & 70784 & 436519 & 336 & 448896 \\
\hline
$c_2^1$-$c_3^3$ & 1268 & 70912 & 439199 & 208 & 263744 \\
\hline
$c_2^2$-$c_3^3$ & 1284 & 72576 & 448871 & 140 & 179760 \\
\hline
$c_2^3$-$c_3^3$ & 1400 & 72576 & 449287 & 140 & 196000 \\
\hline
$c_2^1$-$c_3^4$ & 1716 & 70912 & 442783 & 208 & 356928 \\
\hline
$c_2^2$-$c_3^4$ & 1732 & 72576 & 452455 & 112 & 193984 \\
\hline
$c_2^3$-$c_3^4$ & 1848 & 72576 & 452871 & 112 & 206976 \\
\hline
\end{tabular}
\end{table}

As shown in Tables \ref{tab7-1}, \ref{tab8-1}, and \ref{tab9-1}, the proposed quantum circuits of the S-box significantly reduce the quantum resources needed to implement the quantum circuit for AES. This optimization not only enhances the execution efficiency of the quantum circuit but also provides a feasible solution for implementing larger scale quantum encryption algorithms under limited quantum resources. To compare with other literatures, we select the case with the minimum $T$ depth, and minimum DW ($T$) value. The comparison results are shown in Table \ref{tab10}.

\begin{table}[!t]
\caption{Comparison of the quantum circuits for AES-128.}
\label{tab10}
\centering
\scriptsize
\renewcommand{\arraystretch}{1.3}
\setlength{\tabcolsep}{18pt}
\begin{tabular}{|c|c|c|c|}
\hline
Schemes & \#qubits & $T$ depth & DW($T$) \\
\hline
\text{Ours} & 1028 & 100 & 102800 \\
\hline
\cite{rr42} & 1608 & 80 & 128640 \\
\hline
\cite{rr41} & 3689 & 40 & 147560 \\
\hline
\cite{rr39} & 5576 & 60 & 334560 \\
\hline
\cite{r37} & 256 & 29490 & 7549440 \\
\hline
\end{tabular}
\end{table}

As shown in Table \ref{tab10}, the proposed quantums circuit for AES-128 in this work significantly reduce the DW($T$) value, while maintaining low number of qubits. According to the NIST guidelines for quantum circuit evaluation in Ref.\cite{r29}, the depth and the DW($T$) are considered key metrics for assessing the feasibility of quantum implementations. The depth and width of the quantum circuit are crucial for the performance of practical quantum devices. Compared with Ref.\cite{rr42}, our method achieves a lower DW($T$) value with fewer qubits under the same $T$ depth, reducing it by 20.09\%. Moreover, our method also significantly outperforms Ref.\cite{rr41} and Ref.\cite{rr39} in terms of the DW($T$) value. Compared with Ref.\cite{r37}, although our approach uses more qubits, the DW($T$) value is reduced by 98.64\%. Unlike Refs.\cite{r37,rr39,rr41}, which reduce circuit width at the cost of increased $T$ depth (or vice versa), our AES quantum circuits achieve the lowest DW($T$) value, demonstrating a more efficient depth-width trade-off.

 \section{Conclusion}

This paper presents AES quantum circuits realized with lower DW($T$). We improved upon the method proposed in Ref.\cite{r36} for quantum circuit design, optimizing the AES S-box quantum circuit design by introducing the composite field \( F((2^4)^2) \), which is decomposed into a sequence of operations over the smaller finite field \( F({2^4}) \).

Specifically, we optimize matrix multiplication by reducing the number of $CNOT$ gates, lowering the number of $T$ gates and $T$ depth in the inversion circuit in \( F({2^4}) \), and achieving the quantum multiplication circuit in \( F({2^4}) \) with a reduced width and depth. Combined with the key expansion process, we achieved the lowest known DW($T$) value to date, which is 102800.

\begin{IEEEbiography}[{\includegraphics[width=1in,height=1.25in,clip,keepaspectratio]{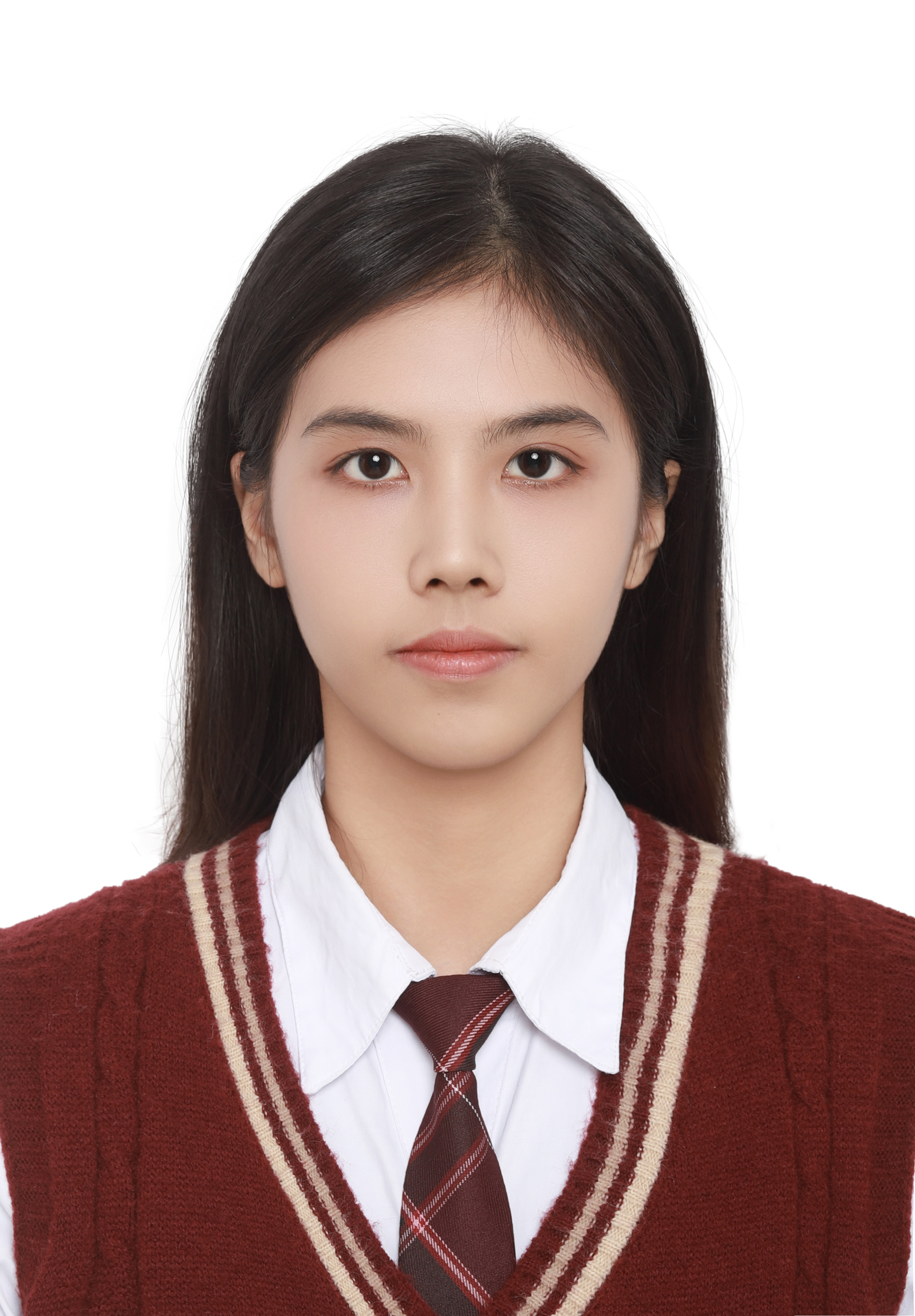}}]{Hui-Nan Chen}
was born in 2001. She is currently pursuing the M.S. degree. Her research interests include quantum cryptography and quantum cryptanalysis.\end{IEEEbiography}

\begin{IEEEbiography}[{\includegraphics[width=1in,height=1.25in,clip,keepaspectratio]{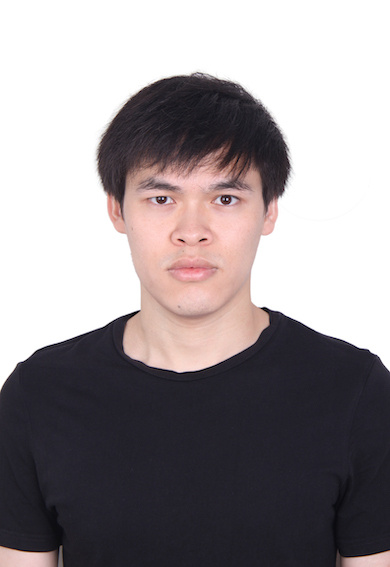}}]{Bin-Bin Cai}
received the B.S. degree in computer science and M.S. degree in computer application both from Fujian Normal University, Fuzhou, China, in 2014 and 2017, respectively, and Ph.D. degree in cyberspace security from Beijing University of Posts and Telcommunications, Beijing, China, in 2023. He is currently a lecturer at the College of Computer and Cyber Security, Fujian Normal University, China. His research interests include quantum cryptography, quantum cryptanalysis and quantum computing.\end{IEEEbiography}

\begin{IEEEbiography}[{\includegraphics[width=1in,height=1.25in,clip,keepaspectratio]{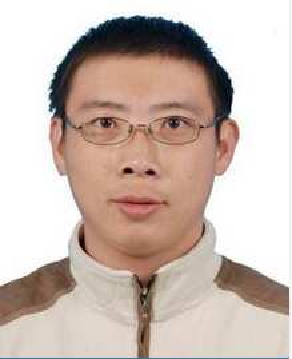}}]{Fei Gao}
received the B.E. degree in communication engineering and the Ph.D. degree in cryptography from the Beijing University of Posts and Telecommunications (BUPT), Beijing, China, in 2002 and 2007, respectively. He is currently the Director of the State Key Laboratory of Networking and Switching Technology, Network Security Research Center (NSRC), BUPT, and also with the Peng Cheng Laboratory, Center for Quantum Computing, Shenzhen, China. He is also working on the practical quantum cryptographic protocols, quantum nonlocality, and quantum algorithms. Prof. Gao is a member of the Chinese Association for Cryptologic Research (CACR).
\end{IEEEbiography}

\begin{IEEEbiography}[{\includegraphics[width=1in,height=1.25in,clip,keepaspectratio]{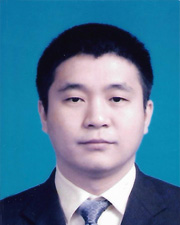}}]{Song Lin}
received the B.S. degree in computer science and M.S. degree in mathematics both from Fujian Normal University, Fuzhou, China, in 1999 and 2005, respectively, and Ph.D. degree in cryptography from Beijing University of Posts and Telcommunications, Beijing, China, in 2009. He is currently a Professor at the College of Computer and Cyber Security, Fujian Normal University, China. His research interests include quantum cryptography, information security, and quantum computing. Prof. Lin is a member of the Chinese Association for Cryptologic Research.\end{IEEEbiography}


\begin{thebibliography}{1}
\bibliographystyle{IEEEtran}
\bibitem{r1}
M. A. Nielsen and I. L. Chuang, \textit{Quantum computation and quantum information}, Cambridge University Press, 2010.

\bibitem{r2}
J. Preskill, ``Quantum computing in the NISQ era and beyond,'' \textit{Quantum}, vol. 2, p. 79, 2018.

\bibitem{r3}
N. S. Yanofsky and M. A. Mannucci, \textit{Quantum computing for computer scientists}, Cambridge University Press, 2008.

\bibitem{r4}
D. Deutsch, ``Quantum theory, the Church-Turing principle and the universal quantum computer,'' \textit{Proc. Roy. Soc. London Ser. A}, vol. 400, pp. 96--117, 1989.

\bibitem{r5}
R. Jozsa, ``Quantum algorithms and the Fourier transform,'' \textit{Proc. Roy. Soc. London Ser. A}, vol. 454, no. 1969, pp. 323--337, 1998.

\bibitem{r6}
A. Ekert and R. Jozsa, ``Quantum computation and Shor's factoring algorithm,'' \textit{Rev. Mod. Phys.}, vol. 68, no. 3, p. 733, 1996.

\bibitem{r7}
A. W. Harrow, A. Hassidim, and S. Lloyd, ``Quantum algorithm for linear systems of equations,'' \textit{Phys. Rev. Lett.}, vol. 103, no. 15, p. 150502, 2009.

\bibitem{r8}
E. Farhi, J. Goldstone, and S. Gutmann, ``A quantum approximate optimization algorithm,'' arXiv preprint arXiv:1411.4028, 2014.

\bibitem{r9}
A. Ajagekar, T. Humble, and F. You, ``Quantum computing based hybrid solution strategies for large-scale discrete-continuous optimization problems,'' \textit{Computers \& Chemical Engineering}, vol. 132, p. 106630, 2020.


\bibitem{r10}
J. J. Su, J. C. Fan, S. Y. Wu, et al., ``Topology-driven quantum architecture search framework,'' \textit{Sci. China Inf. Sci.}, vol. 68, no. 8, p. 180507, 2025.


\bibitem{r11}
S. Y. Wu, Y. Q. Song, R. Z. Li, et al., ``Resource-efficient adaptive variational quantum algorithm for combinatorial optimization problems,'' \textit{Adv. Quantum Technol.}, vol. 2025, p. 2400484, 2025.


\bibitem{r12}
S. Y. Wu, R. Z. Li, Y. Q. Song, et al., ``Quantum assisted hierarchical fuzzy neural network for image classification,'' \textit{IEEE Trans. Fuzzy Syst.}, vol. 33, no. 1, p. 491, 2025.

\bibitem{r13}
P. W. Shor, ``Polynomial-time algorithms for prime factorization and discrete logarithms on a quantum computer,'' \textit{SIAM Review}, vol. 41, no. 2, pp. 303--332, 1999.

\bibitem{r14}
A. Ekert and R. Jozsa, ``Quantum computation and Shor's factoring algorithm,'' \textit{Reviews of Modern Physics}, vol. 68, no. 3, p. 733, 1996.

\bibitem{r15}
L. K. Grover, ``A fast quantum mechanical algorithm for database search,'' in \textit{Proceedings of the Twenty-Eighth Annual ACM Symposium on Theory of Computing}, 1996, pp. 212--219.

\bibitem{r16}
E. Biham, O. Biham, D. Biron, et al., ``Grover＊s quantum search algorithm for an arbitrary initial amplitude distribution,'' \textit{Physical Review A}, vol. 60, no. 4, p. 2742, 1999.

\bibitem{r17}
D. R. Simon, ``On the power of quantum computation,'' \textit{SIAM Journal on Computing}, vol. 26, no. 5, pp. 1474--1483, 1997.

\bibitem{r18}
G. Brassard, P. Hoyer, and A. Tapp, ``Quantum algorithm for the collision problem,'' arXiv preprint quant-ph/9705002, 1997.

\bibitem{r19}
R. Cleve, A. Ekert, C. Macchiavello, et al., ``Quantum algorithms revisited,'' \textit{Proceedings of the Royal Society of London. Series A: Mathematical, Physical and Engineering Sciences}, vol. 454, no. 1969, pp. 339--354, 1998.

\bibitem{r20}
D. Coppersmith, ``An approximate Fourier transform useful in quantum factoring,'' arXiv preprint quant-ph/0201067, 2002.
\bibitem{r21}
A. Y. Kitaev, ``Quantum measurements and the Abelian stabilizer problem,'' arXiv preprint quant-ph/9511026, 1995.

\bibitem{r22}
R. L. Rivest, A. Shamir, and L. Adleman, ``A method for obtaining digital signatures and public-key cryptosystems,'' \textit{Communications of the ACM}, vol. 21, no. 2, pp. 120--126, 1978.

\bibitem{r23}
M. J. Dworkin, E. Barker, J. R. Nechvatal, et al., ``Advanced encryption standard (AES),'' 2001.

\bibitem{r24}
H. Kuwakado and M. Morii, ``Security on the quantum-type Even-Mansour cipher,'' in \textit{2012 International Symposium on Information Theory and Its Applications}, IEEE, 2012, pp. 312--316.

\bibitem{r25}
H. Kuwakado and M. Morii, ``Quantum distinguisher between the 3-round Feistel cipher and the random permutation,'' in \textit{2010 IEEE International Symposium on Information Theory}, IEEE, 2010, pp. 2682--2685.


\bibitem{r42}
S. Jaques, M. Naehrig, M. Roetteler, et al., ``Implementing Grover oracles for quantum key search on AES and LowMC,'' in \textit{Advances in Cryptology每EUROCRYPT 2020: 39th Annual International Conference on the Theory and Applications of Cryptographic Techniques, Zagreb, Croatia, May 10每14, 2020, Proceedings, Part II 30}, Springer International Publishing, 2020, pp. 280--310.

\bibitem{r33}
B. Langenberg, H. Pham, and R. Steinwandt, ``Reducing the cost of implementing the advanced encryption standard as a quantum circuit,'' \textit{IEEE Transactions on Quantum Engineering}, vol. 1, pp. 1--12, 2020.


\bibitem{r43}
M. Amy, D. Maslov, M. Mosca, et al., ``A meet-in-the-middle algorithm for fast synthesis of depth-optimal quantum circuits,'' \textit{IEEE Transactions on Computer-Aided Design of Integrated Circuits and Systems}, vol. 32, no. 6, pp. 818--830, 2013.




\bibitem{r29}
C. F. P. NIST, ``Submission requirements and evaluation criteria for the post-quantum cryptography standardization process,'' 2016.

\bibitem{r30}
L. Chen, L. Chen, S. Jordan, et al., \textit{Report on post-quantum cryptography}, Gaithersburg, MD, USA: US Department of Commerce, National Institute of Standards and Technology, 2016.

\bibitem{r31}
M. Grassl, B. Langenberg, M. Roetteler, et al., ``Applying Grover＊s algorithm to AES: quantum resource estimates,'' in \textit{International Workshop on Post-Quantum Cryptography}, Cham: Springer International Publishing, 2016, pp. 29--43.

\bibitem{r32}
M. Almazrooie, A. Samsudin, R. Abdullah, et al., ``Quantum reversible circuit of AES-128,'' \textit{Quantum Information Processing}, vol. 17, pp. 1--30, 2018.



\bibitem{r34}
J. Zou, Z. Wei, S. Sun, et al., ``Quantum circuit implementations of AES with fewer qubits,'' in \textit{Advances in Cryptology每ASIACRYPT 2020: 26th International Conference on the Theory and Application of Cryptology and Information Security, Daejeon, South Korea, December 7每11, 2020, Proceedings, Part II}, Springer International Publishing, 2020, pp. 697--726.

\bibitem{r35}
Z. G. Wang, S. J. Wei, and G. L. Long, ``A quantum circuit design of AES requiring fewer quantum qubits and gate operations,'' \textit{Frontiers of Physics}, vol. 17, no. 4, p. 41501, 2022.

\bibitem{r36}
Z. Q. Li, B. B. Cai, H. W. Sun, et al., ``Novel quantum circuit implementation of Advanced Encryption Standard with low costs,'' \textit{Science China Physics, Mechanics \& Astronomy}, vol. 65, no. 9, p. 290311, 2022.
\bibitem{rr37}
Z. Li, F. Gao, S. Qin, et al., ``New record in the number of qubits for a quantum implementation of AES,'' \textit{Frontiers in Physics}, vol. 11, p. 1171753, 2023.


\bibitem{r37}
Z. Huang, F. Zhang, and D. Lin, ``Constructing quantum implementations with the minimal T-depth or minimal width and their applications,'' \textit{Cryptology ePrint Archive}, 2025.


\bibitem{rr38}
K. Jang, A. Baksi, H. Kim, et al., ``Improved quantum analysis of SPECK and LowMC,'' in \textit{International Conference on Cryptology in India}, Cham: Springer International Publishing, 2022, pp. 517--540.


\bibitem{rr39}
Z. Huang and S. Sun, ``Synthesizing quantum circuits of AES with lower t-depth and less qubits,'' in \textit{International Conference on the Theory and Application of Cryptology and Information Security}, Cham: Springer Nature Switzerland, 2022, pp. 614--644.

\bibitem{rr40}
J. Boyar and R. Peralta, ``A small depth-16 circuit for the AES S-box,'' in \textit{IFIP International Information Security Conference}, Berlin, Heidelberg: Springer Berlin Heidelberg, 2012, pp. 287--298.



\bibitem{rr41}
Q. Liu, B. Preneel, Z. Zhao, et al., ``Improved quantum circuits for AES: Reducing the depth and the number of qubits,'' in \textit{International Conference on the Theory and Application of Cryptology and Information Security}, Singapore: Springer Nature Singapore, 2023, pp. 67--98.

\bibitem{rr42}
L. L. Jiang, B. B. Cai, F. Gao, et al., ``Constructing resource-efficient quantum circuits for AES,'' \textit{Frontiers in Physics}, vol. 13, p. 1582819, 2025.


\bibitem{r38}
J. Wolkerstorfer, E. Oswald, and M. Lamberger, ``An ASIC implementation of the AES SBoxes,'' in \textit{Topics in Cryptology〞CT-RSA 2002: The Cryptographers＊ Track at the RSA Conference 2002 San Jose, CA, USA, February 18每22, 2002 Proceedings}, Springer Berlin Heidelberg, 2002, pp. 67--78.


\bibitem{r40}
Z. Xiang, X. Zeng, D. Lin, et al., ``Optimizing implementations of linear layers,'' \textit{IACR Transactions on Symmetric Cryptology}, 2020.

\bibitem{r41}
M. Chun, A. Baksi, and A. Chattopadhyay, ``Dorcis: depth optimized quantum implementation of substitution boxes,'' \textit{Cryptology ePrint Archive}, 2023.

\end{thebibliography}
\end{document}